\documentclass[structabstract]{aa}  
\usepackage{textcomp}
\usepackage{graphicx}
\usepackage{txfonts}
\usepackage{natbib}
\usepackage{morefloats}
\usepackage{longtable}
\usepackage{multirow,bigdelim}
\newcommand\nobr{\mbox{-}}
\newcommand\nobrl{\mbox{--}}
\newcommand\kms{km~s$^{-1}$}
\newcommand\jybeam{Jy~beam$^{-1}$}
\begin{document}
   \title{Strong irradiation of protostellar cores in Corona Australis\thanks{Full channel maps (Figs.~\ref{fig:h2co1}--\ref{fig:sio}) of the combined SMA and APEX data are available in electronic form at the CDS via anonymous ftp to \texttt{cdsarc.u-strasbg.fr} (130.79.128.5) or via \texttt{http://cdsweb.u-strasbg.fr/cgi-bin/qcat?J/A+A/}}}

   \author{J. E. Lindberg
          \inst{1,2}
          \and
          J. K. J{\o}rgensen\inst{2,1}
          }

   \institute{{Centre for Star and Planet Formation, Natural History Museum of Denmark, University of Copenhagen, {\O}ster Voldgade 5-7, DK\nobr1350 K{\o}benhavn K, Denmark}\\
              \email{jlindberg@snm.ku.dk}
         	\and
      {Niels Bohr Institute, University of Copenhagen, Juliane Maries Vej 30, DK\nobr2100 K{\o}benhavn {\O}, Denmark}\\          
         }

   \date{Received May 15, 2012; accepted September 19, 2012}

 
  \abstract
   {The importance of the physical environment in the evolution of newly formed low-mass stars remains an open question. In particular, radiation from nearby more massive stars may affect both the physical and chemical structure of these kinds of young stars.}
   {To constrain the physical characteristics of a group of embedded low-mass protostars in Corona Australis in the vicinity of the young luminous Herbig Be star R~CrA.}
   {Millimetre wavelength maps of molecular line and continuum emission towards the low-mass star forming region IRS7 near R~CrA from the Submillimeter Array (SMA) and Atacama Pathfinder Experiment (APEX) are presented. The maps show the distribution of 18~lines from 7~species (H$_2$CO, CH$_3$OH, HC$_3$N, \textit{c}\nobr C$_3$H$_2$, HCN, CN and SiO) on scales from 3\arcsec\ to 60\arcsec\ (400--8\,000~AU). Using a set of H$_2$CO lines, we estimate the temperatures and column densities in the region using both LTE and non-LTE methods. The results are compared with 1-D radiative transfer modelling of the protostellar cores. These models constrain which properties of the central source, protostellar envelope, and surrounding radiation field can give rise to the observed line and continuum emission.}
   {Most of the H$_2$CO emission from the regions emerges from two elongated ($\sim 6\,000$~AU long) narrow ($< 1\,500 $~AU) ridges dominating the emission picked up in both interferometric and single-dish measurements. The temperatures inferred from the H$_2$CO lines are no less than $\sim30$~K and more likely 50--60~K, and the line emission peaks are offset by $\sim2\,500$~AU from the location of the embedded protostars. These temperatures can not be explained by the heating from the young stellar objects (YSOs) themselves. Irradiation by the nearby Herbig~Be star R~CrA could, however, explain these high temperatures. The elevated temperatures can in turn impact the physical and chemical characteristics of protostars, in particular, lead to enhanced abundances of typical tracers of photon dominated regions (PDRs) such as seen in single-dish line surveys of embedded protostars in the region.}
   {}

   \titlerunning{Strong irradiation of protostellar cores in CrA}
   \maketitle
%

\section{Introduction}
\label{sec:intro}

Low-mass stars form when dense molecular cloud cores undergo gravitational collapse. In the very early stages \citep[Class~0;][]{andre93,andre00}, the young star is obscured by its envelope of gas and dust. This envelope eventually dissipates, and leaves a pre-main-sequence star with a planet-forming disc. One of the interesting questions concerning the star formation process is how important the ambient environment is for the outcome, e.g. in terms of the physical and chemical properties of the emerging protostars and their discs. It is for example well-known that high-mass stars in the environment of young stellar objects (YSOs) can affect both the chemical and physical evolution of the protostars \citep[see e.g.][]{tielens85a,tielens85b}. The most well-known examples of young objects affected by highly luminous stars are the proplyds in the Orion Nebula, which were discovered in \textit{Hubble Space Telescope} observations \citep{odell93}. In these sources, the luminous emission from the high-mass stars results in photoevaporation of the protoplanetary discs \citep{johnstone98}. In this paper, we focus on the contribution of external irradiation onto more deeply embedded sources in the Corona Australis (CrA) region.

The CrA dark cloud is one of the most nearby active star forming regions at a distance of 130~pc \citep{neuhauser08}. Through multi-wavelength studies of the CrA region, \citet{peterson11} found 116 YSO candidates, the vast majority being Class~II and Class~III sources. The region around the Herbig Be star R~CrA -- which is also referred to as the Coronet cluster -- contains the highest concentration of newly formed and young stars \citep[Class~I and earlier;][]{peterson11}. R~CrA itself is a Herbig Be star \citep[spectral class B5--B8;][]{gray06,bibo92} with a mass of 3.0~$M_{\sun}$ \citep{bibo92} and a luminosity between 100 and 166~$L_{\sun}$ \citep[the lower value is a lower limit from SED estimates with incomplete data in the UV range;][]{bibo92}. Several dense molecular cloud cores with masses between 2 and 50~$M_{\sun}$ were found near the Herbig Be star R~CrA in C$^{18}$O SEST observations by \citet{harju93}. The IRS7 region is located with its centre between the Herbig Be star R~CrA and the T~Tauri star T~CrA \citep{taylor84}, $\sim$30\arcsec\ ($\approx$ 4\,000~AU) from each of these stars, and harbours a handful of Class~0/I YSOs.

\begin{figure*}[!htb]
	\centering
    \includegraphics{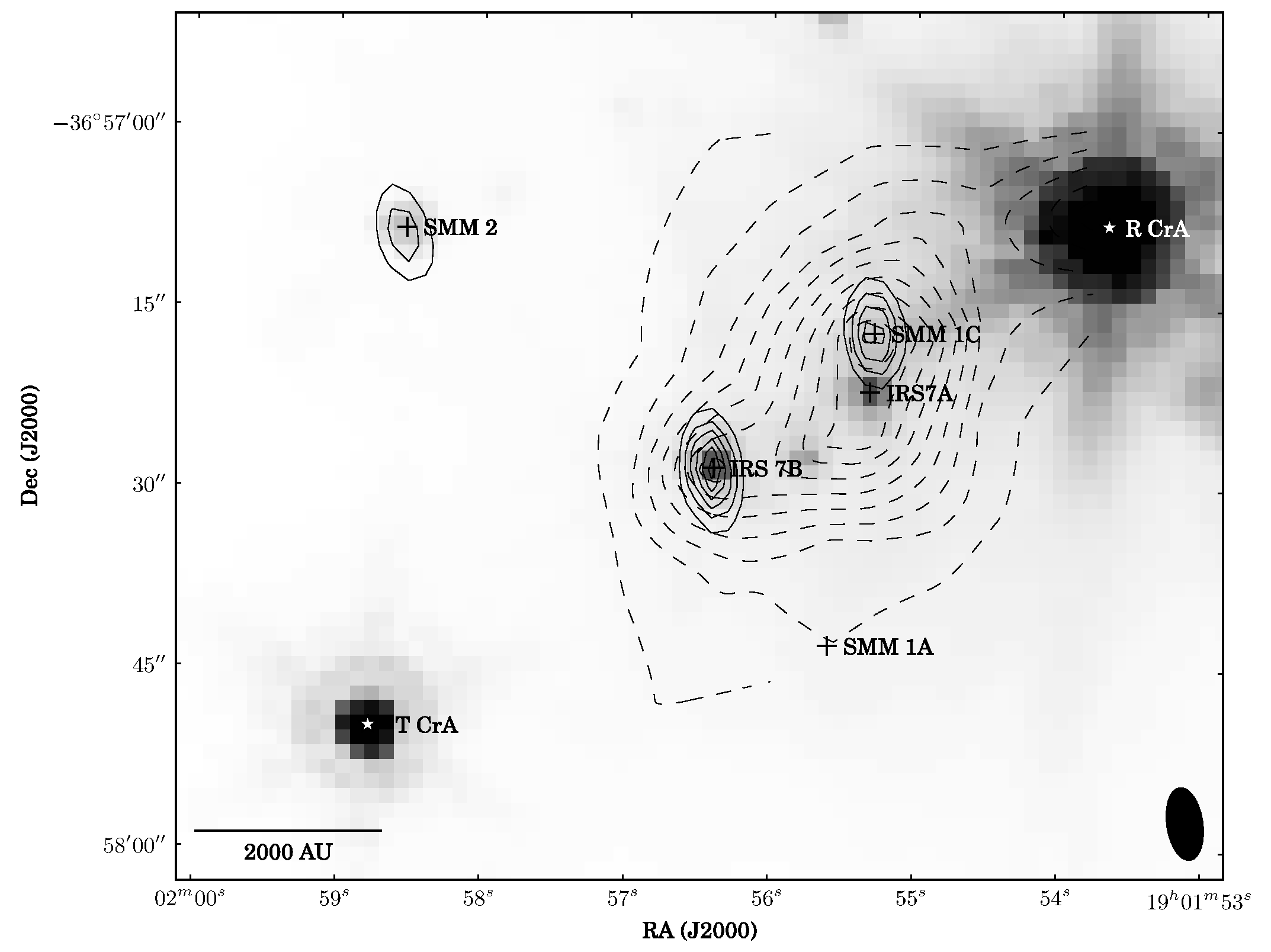}
    \caption{SMA 1.3~mm continuum (10$\sigma$ solid contours, $\sigma = 4.7$ m\jybeam), \textit{Herschel} PACS 70~\hbox{\textmu}m continuum (10~Jy dashed contours; see also Lindberg et~al., in prep.), and \textit{Spitzer} 4.5~\hbox{\textmu}m image \citep[greyscale; see also][]{peterson11}. The \textit{Herschel} PACS beam size is 9\arcsec, and the data consist of two overlapping $5\times5$ spaxel arrays, with a $9.4$\arcsec$\times9.4$\arcsec spaxel size. The SMA beam size ($6.0$\arcsec$\times2.9$\arcsec) is shown in the bottom right corner. The plus signs indicate the positions of the YSOs \citep{peterson11} and SMM~1A \citep{nutter05}, and the star symbols indicate the positions of R~CrA and T~CrA \citep{peterson11}. These symbols will be used throughout the paper.}
    \label{fig:herschelsmaspitzer}
\end{figure*}

A summary of continuum point sources in the region around R~CrA reported in the literature is given in Table~\ref{tab:pointsource}. The locations of the sources are shown on \textit{Spitzer} 4.5~\hbox{\textmu}m \citep{peterson11} image data, with \textit{Herschel} 70~\hbox{\textmu}m (Lindberg et al., in prep.) and SMA 1.3~mm \citep{chen10} continuum data overplotted in Fig.~\ref{fig:herschelsmaspitzer}. Using NIR photometry, \citet{taylor84} found that IRS7 is the most reddened source in the R~CrA region, with a visual extinction of more than 25~mag. Through VLA 6~cm observations, \citet{brown87} discovered that IRS7 is of binary nature, and separated it into IRS7A and IRS7B, two apparently compact sources with a separation of 10--15\arcsec\ (2\,000~AU). \citet{brown87} also detected a faint continuum source north of IRS7A. \citet{miettinen08} referred to this as Brown~9, and used 3~cm and 6~cm continuum observations to produce high-resolution high-sensitivity maps of the region. \citet{nutter05} studied the region in submillimetre, and detected three peaks within the region around IRS7: SMM~1B coinciding with IRS7B, SMM~1C coinciding with Brown~9, and SMM~1A south of IRS7B. Studies of millimetre continuum emission showed three compact sources, IRS7B (SMM~1B), SMM~1C, and SMM~2 \citep{groppi07,chen10,peterson11}. \citet{groppi07} used \textit{Spitzer} IRAC data, SCUBA 450~\hbox{\textmu}m and 850~\hbox{\textmu}m, and SMA 1.1~mm data to study the SEDs of IRS7B (SMA1) and SMM~1C (SMA2; not detected in the \textit{Spitzer} data). They suggested that IRS7B is a transitional Class~0/I object, whereas SMM~1C likely is a Class~0 source. IRS7A shows strong mid-IR emission (\textit{Spitzer}), but is not detected in the (sub-)millimetre. Based on their IR and sub-millimetre SEDs, \citet{peterson11} suggested that IRS7A and SMM~2 are Class~I young stellar objects. SMM~1A is suggested to be a prestellar core \citep{chen10}. 

Compared to other typical embedded protostars IRS7B has a fairly rich line spectrum and is of interest as a possible astrochemical laboratory. Detailed studies of the chemistry of IRS7B have been made possible with a new generation of southern hemisphere submillimetre telescopes such as Atacama Pathfinder Experiment (APEX) \citep{schoier06} and Atacama Submillimeter Telescope Experiment (ASTE) \citep{watanabe12}. \citeauthor{watanabe12} performed a full line scan of IRS7B in the 345~GHz submillimetre window using ASTE, and found H$_2$CO and CH$_3$OH, two molecules typical for a chemistry dominated by evaporation of icy grain mantles at temperatures higher than 90--100~K \citep[see e.g.][]{bottinelli04a,bottinelli04b,ceccarelli07}, but also species indicative of a warm carbon-chain chemistry \citep[CCH and \textit{c}\nobr C$_3$H$_2$; ][]{sakai09a,sakai09b}. \citeauthor{watanabe12} suggest that IRS7B could be an intermediate-type source between the typical ``hot corinos'' and ``warm carbon-chain chemistry sources''. However, a different possibility could be that the observed chemical composition is a result of photodissociation caused by bright FUV radiation from the nearby star R~CrA further underscoring the importance of understanding the properties of the protostar's environment.

In this paper we present high angular resolution millimetre maps of the line emission in the region surrounding R~CrA IRS7 and aim to quantify the possible physical and chemical impact of the irradiation by the Herbig Be star R~CrA. The paper is laid out as follows: Sect.~\ref{sec:observations} describes the details of the observations and reduction techniques. Sect.~\ref{sec:results} shows the results of these observations. These results are then analysed with LTE and non-LTE modelling in Sect.~\ref{sec:analysis}. Finally, Sect.~\ref{sec:discussion} discusses the implications of the derived physical and chemical properties.

\begin{table*}[!htb]
\centering
\caption[]{Continuum point sources in the SMA, \textit{Spitzer}, and SCUBA data.}
\label{tab:pointsource}
\begin{tabular}{l l l l l}
\noalign{\smallskip}
\hline
\hline
\noalign{\smallskip}
Source name & RA & Dec & Other names\tablefootmark{a} & Coordinate reference\tablefootmark{b} \\
& (J2000.0) & (J2000.0) & \\
\noalign{\smallskip}
\hline
\noalign{\smallskip}
SMM~1C & 19:01:55.29 & $-$36:57:17.0 & SMA~2 (1,2), B9 (3), Brown~9 (4) & SMA \\
IRS7A & 19:01:55.32 & $-$36:57:21.9 & IRS7W (5), IRS7 (2,6) & \textit{Spitzer} \\
IRS7B & 19:01:56.40 & $-$36:57:28.3 & SMM 1B (6,7), SMA~1 (2), IRS7E (5) & SMA \\
SMM~2 & 19:01:58.54 & $-$36:57:08.5 & CrA-43 (1), WMB 55 (3,6) & \textit{Spitzer} \\
R~CrA & 19:01:53.67 & $-$36:57:08.0 & -- & \textit{Spitzer} \\
\noalign{\smallskip}
\hline
\end{tabular}
\tablefoot{
     	\tablefoottext{a}{References: (1)~\citet{peterson11}, (2)~\citet{groppi07}, (3)~\citet{choi08}, (4)~\citet{miettinen08}, (5)~\citet{forbrich06}, (6)~\citet{nutter05}, (7)~\citet{chen10}.}\\
     	\tablefoottext{b}{The coordinates come from SMA and \textit{Spitzer} continuum reported in \citet{peterson11}.}\\
     	}
\end{table*}

\section{Observations}
\label{sec:observations}

\subsection{SMA 1.3 mm observations}
\begin{figure}[!htb]
	\centering
    \includegraphics{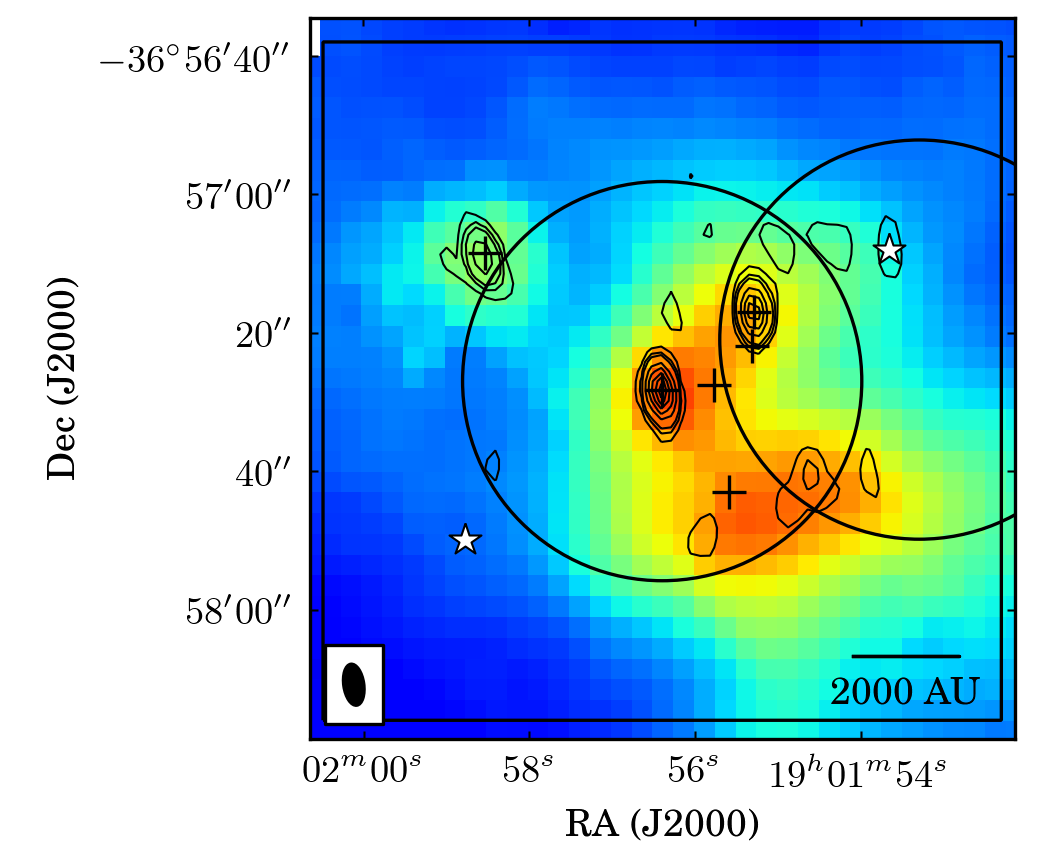}
    \caption{SMA 1.3~mm continuum (contours at 3$\sigma$, 6$\sigma$, 9$\sigma$, and then increasing in steps of 10$\sigma$, where $\sigma = 4.7$ m\jybeam) and SCUBA 850~\hbox{\textmu}m continuum (colour scale). The two circles show the primary beams of the two SMA pointings. The box shows the coverage of the APEX short-spacing data. The synthetic beam size of the SMA observations is $6.0$\arcsec$\times2.9$\arcsec. Refer to Fig.\ref{fig:herschelsmaspitzer} for a guide to the symbols used for the compact objects.}
    \label{fig:smascuba}
\end{figure}

The region around IRS7A, IRS7B (SMM~1B), SMM~1C, SMM~1A, and R~CrA was observed at 1.3~mm using the Submillimeter Array (SMA; \citealt{ho04})\footnote{The Submillimeter Array is a joint project between the Smithsonian Astrophysical Observatory and the Academia Sinica Institute of Astronomy and Astrophysics and is funded by the Smithsonian Institution and the Academia Sinica.} in its compact configuration on 20 August 2006. Six antennae were in the array covering baselines from 5--50~k$\lambda$. The region was covered by two pointings, one located on IRS7B, and one south of R~CrA. Each of the two pointings covers a circular area with a diameter of about 7\,500~AU dictated by the SMA primary beam size of 58\arcsec, but the two pointings overlap somewhat. The exact coordinates of the pointings can be found in Table~\ref{tab:obsparam}. The SMA 1.3~mm continuum is overplotted on the SCUBA 850~\hbox{\textmu}m continuum \citep{nutter05} together with the SMA primary beams at the wavelength of the observations in Fig.~\ref{fig:smascuba}.

Both pointings were observed with the lower sideband covering 216.849--218.831~GHz, and the upper sideband covering 226.849--228.831~GHz. The SMA correlator was configured to give uniform spectral coverage over each of these 2~GHz bandwidths. The resulting frequency resolution of 0.81~MHz\ corresponds to a velocity resolution of 1.1~km~s$^{-1}$ at the observed frequencies. The data were calibrated using the MIR package \citep{qimir} dedicated to SMA observations. The complex gains were calibrated through observations of the quasar 1924--292 with the bandpass and absolute levels calibrated through observations of the strong quasar 3c454.3 and Uranus. The resulting synthetic beam size of the observations was $6.0$\arcsec$\times2.9$\arcsec\ with a position angle of 5.3\degr.

Further imaging of the data was performed with the MIRIAD package \citep{sault95}. First, a linear fit to the continuum was subtracted from the $(u,v)$ files. The sky brightness distribution was then produced from an inverse Fourier transform of the $(u,v)$ data sets, and consecutively CLEANed. Finally, the data from the two separate pointings were combined by use of the \texttt{linmos} (linear mosaic) task. The 1.3~mm continuum data were produced by averaging the $(u,v)$ data across both SMA sidebands. The contribution from the strong line emission was estimated to be less than $10\%$ in the observed continuum peaks (including SMM~1A). Previous papers by \cite{chen10} and \cite{peterson11} have discussed the continuum data from these observations separately; here we focus predominantly on the line observations.

The SMA line data were subsequently combined with APEX short-spacing data (see below).

\begin{table}
\centering
\caption[]{Pointing centres and frequency coverage for the SMA and APEX observations.}
\label{tab:obsparam}
\begin{tabular}{l l l l}
\noalign{\smallskip}
\hline
\hline
\noalign{\smallskip}
\textit{Telescope} & R.A. & Dec. & Line data \\ 
\hspace{3mm}Region & (J2000.0) & (J2000.0) & [GHz] \\ 
\noalign{\smallskip}
\hline
\noalign{\smallskip}
\textit{SMA} & & & \\
\hspace{3mm}IRS7A & 19:01:53.30 & $-$36:57:21.0 & 217.8$\pm$1.0 \\
\hspace{3mm}IRS7A & 19:01:53.30 & $-$36:57:21.0 & 227.8$\pm$1.0 \\
\hspace{3mm}IRS7B & 19:01:56.40 & $-$36:57:27.0 & 217.8$\pm$1.0 \\
\hspace{3mm}IRS7B & 19:01:56.40 & $-$36:57:27.0 & 227.8$\pm$1.0 \\
\noalign{\smallskip}
\hline
\noalign{\smallskip}
\textit{APEX} & & & \\
\hspace{3mm}IRS7B\tablefootmark{a} & 19:01:56.40 & $-$36:57:27.0 & 217.9$\pm$0.9 \\
\noalign{\smallskip}
\hline
\end{tabular}
\tablefoot{
     	\tablefoottext{a}{$7\times7$ pointings in an equatorial grid centred at the tabulated coordinate. The separation between the grid points is 14\arcsec.} \\
    	}
\end{table}

\subsection{APEX 1.3 mm observations}
\label{sec:apexobs}

To study the extended molecular emission in the region, short-spacing data for the SMA observations were obtained with APEX\footnote{The Atacama Pathfinder EXperiment (APEX) telescope is a collaboration between the Max Planck Institute for Radio Astronomy, the European Southern Observatory, and the Onsala Space Observatory}. The region was observed in a $7 \times 7$ pointing pattern centred at IRS7B, covering most of the SMA primary beams (Fig.\ref{fig:smascuba}; the westernmost part of the region covered by SMA was not covered in the APEX map, since no signal was detected in the SMA maps at those positions) in the frequency range between 216.951~GHz and 218.849~GHz (corresponding to the SMA lower sideband). The beam size of the APEX telescope at these frequencies is 28.6\arcsec, and the individual pointings were therefore separated by 14\arcsec\ to give a fully Nyquist-sampled map. The frequency resolution of the APEX observations (0.12~MHz) corresponds to a velocity resolution of 0.17~\kms.

The APEX data were reduced with the GILDAS CLASS package\footnote{GILDAS CLASS (Continuum and Line Analysis Single-dish Software) is developed by the IRAM institute, Grenoble, France: \texttt{http://www.iram.fr/IRAMFR/GILDAS}}. Sinusoidal or, in some cases, polynomial ($n\leq5$), baselines were carefully calculated and subtracted. The data were then exported to FITS format, to allow for import of the data to MIRIAD.

The APEX data were smoothed to match the velocity resolution of the SMA data with the \texttt{regrid} task in MIRIAD. The two data sets were then combined with the MIRIAD \texttt{immerge} task, which linearly merges two sets of data with different spatial resolution (i.e. applies short-spacing corrections from the single-dish data onto the interferometry data). In the combination, primary beam correction was applied, which amplifies the noise towards the edges of the primary beam. The estimated errors in figures in this paper are calculated at the edges of the primary beams. However, in the estimates of fluxes in individual positions in the maps, the RMS noise in that particular point has been used for the error analysis. 

This linear method (\texttt{immerge} in MIRIAD) of combining single-dish and interferometry data was chosen in favour of the non-linear alternative (\texttt{mosmem} in MIRIAD). The reason for this is that the separation between the two SMA pointings is greater than 0.5 times the primary beam FWHM, making the mosaic somewhat undersampled. The non-linear method would thus risk to amplify noise, in particular near the primary beam edges. A comparison between the two methods also showed this, with flux differences of less than $15\%$ in the inner parts of the primary beam, but the non-linear method over-estimating the signal close to the primary beam edges.

\subsection{\textit{Spitzer}, JCMT/SCUBA, and \textit{Herschel} data}

In addition to the described SMA and APEX observations we utilise continuum maps from mid- and far-infrared space-based observations from the \textit{Spitzer Space Telescope} \citep{peterson11} and \textit{Herschel Space Observatory} (Lindberg et~al., in prep.) as well as ground-based submillimetre observations from the SCUBA camera at the \textit{James Clerk Maxwell Telescope} \citep{nutter05}.

\section{Results}
\label{sec:results}

Eighteen spectral lines from seven molecular species were detected in the 216.8--218.8~GHz and 226.8--228.8~GHz SMA data. The detected species are H$_2$CO, CH$_3$OH, \textit{c}\nobr C$_3$H$_2$, HC$_3$N, DCN, CN (both $^{12}$CN and $^{13}$CN), and SiO (see Table~\ref{tab:molsma}). Most of these lines, in particular the H$_2$CO and CH$_3$OH lines, show very extended emission, which in most cases does not coincide with any of the bright compact continuum sources associated with the main YSOs, but does align with e.g. the SCUBA peak SMM~1A, south of the suggested YSOs and is also detected in a region between R~CrA and IRS7.\footnote{Note that \citet{chen10} report that no molecular line emission from SMM~1A is seen in the same data; contrasting what is seen in the results presented here from our reduction of the data.}

\begin{table}
\centering
\caption[]{Identified molecular lines in the SMA/APEX data.}
\label{tab:molsma}
\begin{tabular}{l l l @{}l r}
\noalign{\smallskip}
\hline
\hline
\noalign{\smallskip}
Species & Transition & & Frequency\tablefootmark{a} & $E_\mathrm{u}$\tablefootmark{a} \\
&  & & [GHz] & [K] \\
\noalign{\smallskip}
\hline
\noalign{\smallskip}
SiO & $J$ = 5\nobrl 4 & & 217.10498 & 31.26 \\
DCN & $J$ = 3\nobrl 2 & & 217.23854 & 20.85 \\
$^{13}$CN & $N$ = 2\nobrl 1, $J$ = 3/2\nobrl 1/2\tablefootmark{b} & & 217.26464 & 15.66 \\
$^{13}$CN & $N$ = 2\nobrl 1, $J$ = 5/2\nobrl 3/2\tablefootmark{c} & & 217.46715 & 15.68 \\
\textit{c}\nobr C$_3$H$_2$\tablefootmark{d} & 6$_{16}$\nobrl 5$_{05}$ & \ldelim\{{2}{2.5mm}[] & 217.82215 & 38.61 \\
\textit{c}\nobr C$_3$H$_2$\tablefootmark{d} & 6$_{06}$\nobrl 5$_{15}$ & & 217.82215 & 38.61 \\
\textit{c}\nobr C$_3$H$_2$ & 5$_{14}$\nobrl 4$_{23}$ & & 217.94005 & 35.42 \\
\textit{c}\nobr C$_3$H$_2$ & 5$_{24}$\nobrl 4$_{13}$ & & 218.16044 & 35.42 \\
H$_2$CO & 3$_{03}$\nobrl 2$_{02}$ & & 218.22219 & 20.96 \\
HC$_3$N & $J$ = 24\nobrl 23 & & 218.32472 & 130.98 \\
CH$_3$OH & 4$_{22}$\nobrl 3$_{12}$ & & 218.44005 & 31.60 \\
H$_2$CO & 3$_{22}$\nobrl 2$_{21}$ & & 218.47563 & 68.09 \\
H$_2$CO & 3$_{21}$\nobrl 2$_{20}$ & & 218.76007 & 68.11 \\
$^{12}$CN\tablefootmark{e} & $N$ = 2\nobrl 1, $J$ = 5/2\nobrl 3/2\tablefootmark{f} & & 226.87478 & 11.35 \\
$^{12}$CN\tablefootmark{e} & $N$ = 2\nobrl 1, $J$ = 5/2\nobrl 3/2\tablefootmark{g} & & 226.88742 & 11.35 \\
$^{12}$CN\tablefootmark{e} & $N$ = 2\nobrl 1, $J$ = 5/2\nobrl 3/2\tablefootmark{h} & & 226.89213 & 11.35 \\
\textit{c}\nobr C$_3$H$_2$\tablefootmark{e} & 4$_{32}$\nobrl 3$_{21}$ & & 227.16913 & 29.07 \\
HC$_3$N\tablefootmark{e} & $J$ = 25\nobrl 24 & & 227.41890 & 141.90 \\
\noalign{\smallskip}
\hline
\end{tabular}
\tablefoot{
     	\tablefoottext{a}{Frequencies and energies were acquired from the CDMS database \citep{cdms}, except for the \textit{c}\nobr C$_3$H$_2$ data, which are from the JPL database \citep{jpl}.} \\
		\tablefoottext{b}{$F_1$ = 1\nobrl 0, $F$ = 0\nobrl 1}
		\tablefoottext{c}{Blend of $F_1$ = 3\nobrl 2, $F$ = 4\nobrl 3; $F_1$ = 3\nobrl 2, $F$ = 3\nobrl 2; and  $F_1$ = 3\nobrl 2, $F$ = 2\nobrl 1.}
		\tablefoottext{d}{This is a blend of two \textit{c}\nobr C$_3$H$_2$ lines.}\\
     	\tablefoottext{e}{Only SMA data are available for these lines.}
     	\tablefoottext{f}{Blend of $F$ = 5/2\nobrl 3/2,  $F$ = 7/2\nobrl 5/2, and $F$ = 3/2\nobrl 1/2.}
     	\tablefoottext{g}{$F$ = 3/2\nobrl 3/2.}
     	\tablefoottext{h}{$F$ = 5/2\nobrl 5/2.}
     	}
\end{table}

\begin{figure*}[!htb]
    \centering
    \includegraphics{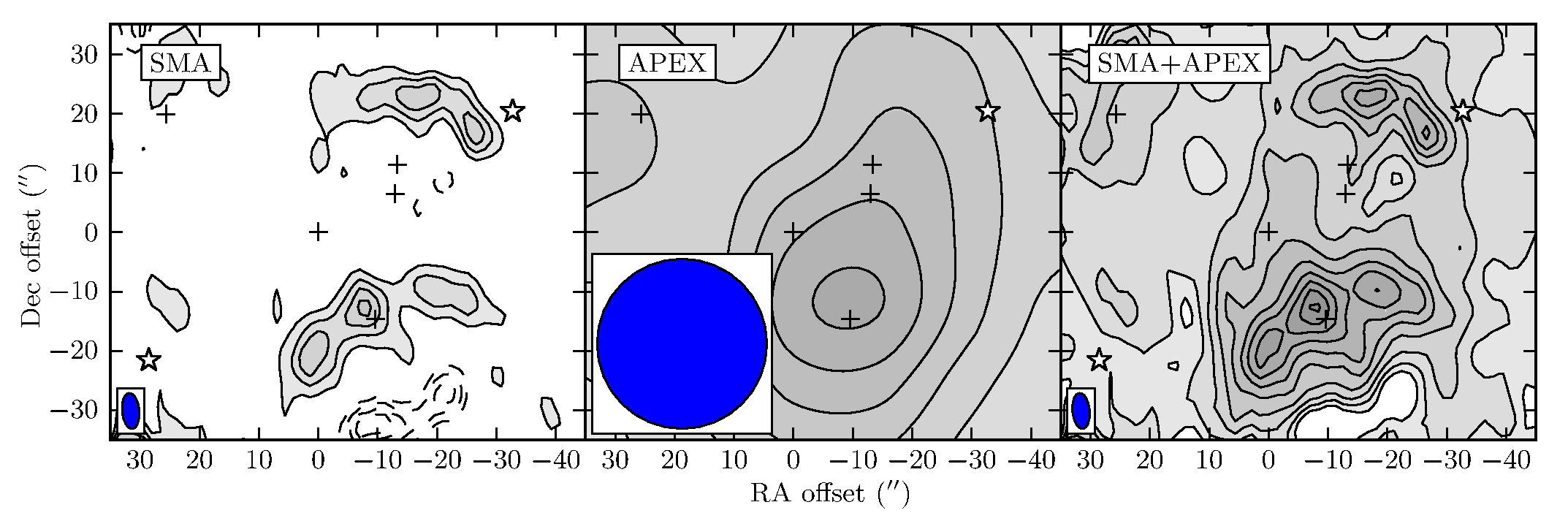} \\
    \caption{\textit{Left:} H$_2$CO 3$_{03}$\nobrl 2$_{02}$ line emission detected by the SMA integrated between 0 and +9~\kms. \textit{Middle:} H$_2$CO 3$_{03}$\nobrl 2$_{02}$ line emission observed by the APEX telescope in a 7$\times$7 grid integrated between 0 and +9~\kms. \textit{Right:} H$_2$CO 3$_{03}$\nobrl 2$_{02}$ line emission in the combined SMA interferometry and APEX short-spacing data integrated between 0 and +9~\kms. \newline Contours are at 2.5~\jybeam~\kms\ levels (converted to SMA beams, $\sim 5\sigma$ at the edge of the SMA primary beam) in all three maps. Negative intensities are represented by dashed contours. Coordinates are offsets from the central position of IRS7B. Refer to Fig.\ref{fig:herschelsmaspitzer} for a guide to the symbols used for the compact objects.}
     \label{fig:short_compare1}
\end{figure*}
\begin{figure}[!htb]
	\centering
	\includegraphics{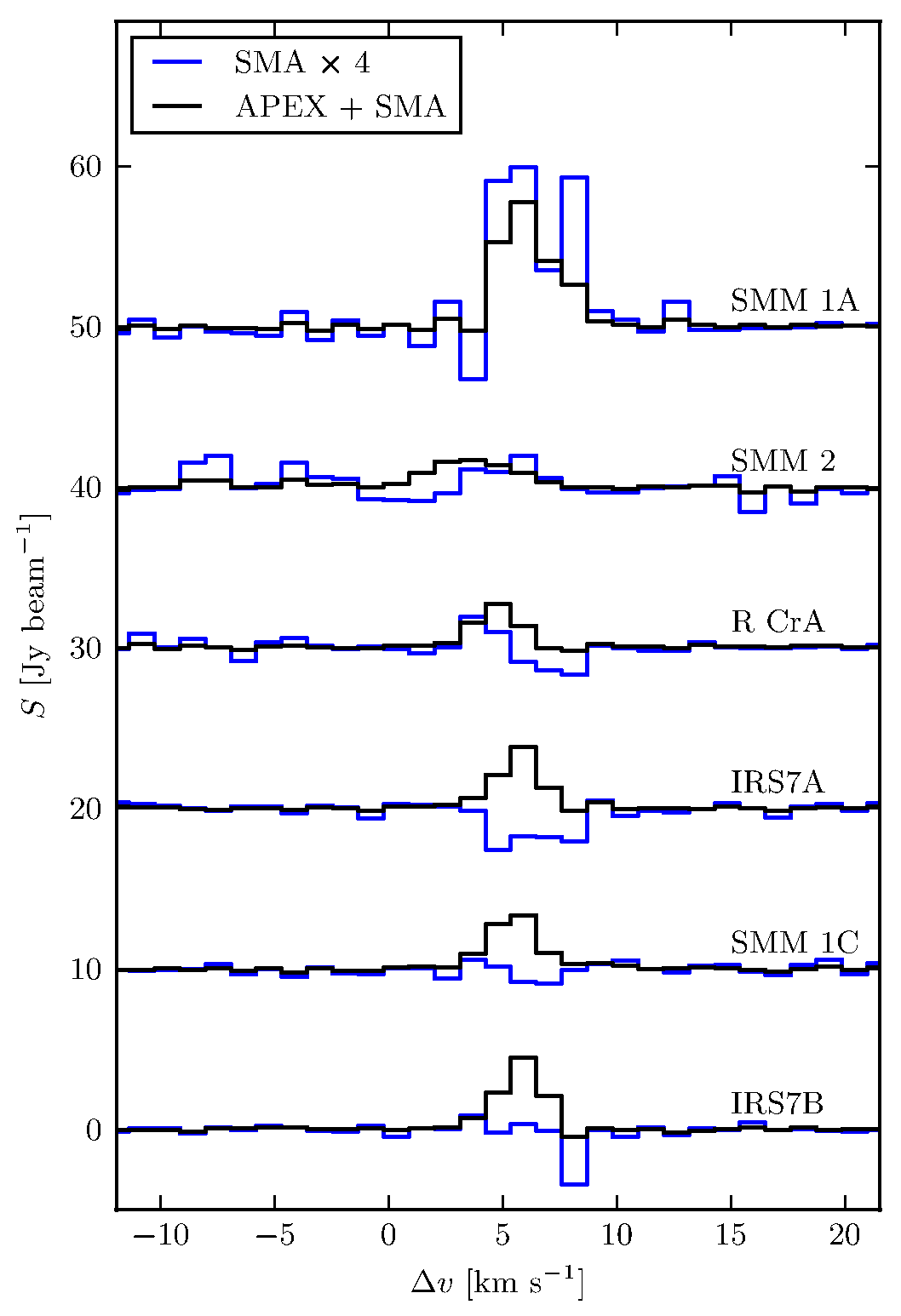}
	\caption{Comparison between SMA (blue) and SMA/APEX (black) H$_2$CO 3$_{03}$\nobrl 2$_{02}$ spectra in SMM~1A, SMM~2, R~CrA, IRS7A, SMM~1C, and IRS7B. The SMA spectra have been scaled by a factor of 4, and there is a 10~\jybeam\ vertical shift between the spectra. Units are in \jybeam\ for the SMA beam ($6.0$\arcsec$\times2.9$\arcsec).}
	\label{fig:short_compare2}
\end{figure}

In Fig.~\ref{fig:short_compare1}, the integrated intensity maps of the H$_2$CO 3$_{03}$\nobrl 2$_{02}$ line using SMA data, APEX data, and the combined data set are compared. The bulk of the emission is extended and thus completely resolved-out in the pure interferometry data. Combining the SMA data with the large-scale APEX maps is therefore critical in order to derive quantitative numbers from the data set. All 12 spectral lines detected in the SMA 216.8--218.8~GHz band were also detected in the APEX data. The SMA and APEX data were combined following the methods described in \citet{weiss01} and \citet{takakuwa03} (see also Sect.~\ref{sec:apexobs}). In Fig.~\ref{fig:short_compare2}, the pure interferometry fluxes and the short-spacing-corrected fluxes of the H$_2$CO 3$_{03}$\nobrl 2$_{02}$ line from six positions in the interferometry map are compared. The pure interferometry lines are not only fainter, they also show absorption in several positions, since the real fluxes in those positions are lower than the average of the extended flux in the field. Typically, less than $25\%$ of the peak flux is recovered in the interferometry data. 

\begin{figure*}[!htb]
    \centering    
	\includegraphics{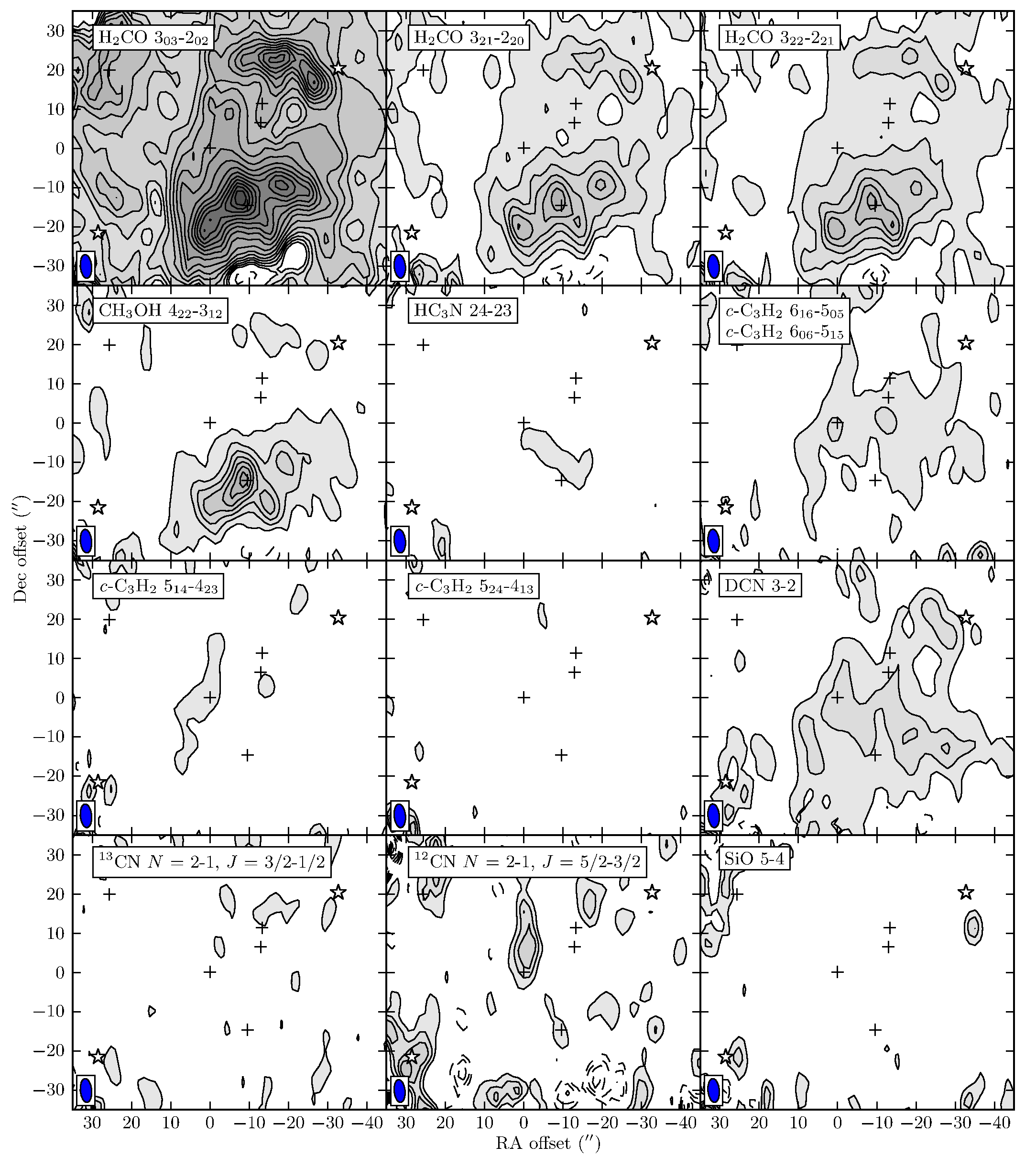} \\
    \caption{SMA/APEX integrated intensity line maps of all spectral lines in the SMA coverage except for $^{13}$CN $N$ = 2\nobrl 1, $J$ = 5/2\nobrl 3/2 (blend of several hyperfine transitions), which is barely detected in the interferometry data. The plotted $^{13}$CN line has the quantum numbers $F_1$ = 1\nobrl 0, $F$ = 0\nobrl 1. No APEX short-spacing data were used for the  $^{12}$CN $N$ = 2\nobrl 1, $J$ = 5/2\nobrl 3/2 (blend of several hyperfine transitions) line. All contour levels are at a $3\sigma$ level at the edge of the primary beam. Refer to Table~\ref{tab:figprops} for the exact contour levels and the integration intervals. Refer to Fig.\ref{fig:herschelsmaspitzer} for a guide to the symbols used for the compact objects.}
    \label{fig:int1}
\end{figure*}

\begin{table}[!htb]
\centering
\caption[]{Integration intervals and contour levels of the integrated intensity maps in Fig.~\ref{fig:int1}. All contour levels are at a $3\sigma$ level at the edge of the primary beam.}
\label{tab:figprops}
\begin{tabular}{l c c}
\noalign{\smallskip}
\hline
\hline
\noalign{\smallskip}
Spectral line & $v$ & Contour level \\
& [\kms] & [\jybeam \kms] \\
\noalign{\smallskip}
\hline
\noalign{\smallskip}
H$_2$CO 3$_{03}$\nobrl 2$_{02}$ & $[-0.2;+8.7]$ & 1.46 \\
H$_2$CO 3$_{21}$\nobrl 2$_{20}$ & $[+3.0;+7.5]$ & 1.03 \\
H$_2$CO 3$_{22}$\nobrl 2$_{21}$ & $[+2.6;+8.2]$ & 1.15 \\
CH$_3$OH 4$_{22}$\nobrl 3$_{12}$ & $[+2.8;+7.3]$ & 1.03 \\
HC$_3$N $J$ = 24\nobrl 23 & $[+4.5;+7.9]$ & 0.89 \\
\textit{c}\nobr C$_3$H$_2$ 6$_{16}$\nobrl 5$_{05}$, 6$_{06}$\nobrl 5$_{15}$ & $[+3.7;+7.1]$ & 0.89 \\
\textit{c}\nobr C$_3$H$_2$ 5$_{14}$\nobrl 4$_{23}$ & $[+4.8;+7.0]$ & 0.73 \\
\textit{c}\nobr C$_3$H$_2$ 5$_{24}$\nobrl 4$_{13}$ & $[+3.5;+5.7]$ & 0.73 \\
$^{13}$CN $N$ = 2\nobrl 1, $J$ = 3/2\nobrl 1/2\tablefootmark{a} & $[+4.9;+8.3]$ & 0.90 \\
DCN $J$ = 3\nobrl 2 & $[+4.8;+7.0]$ & 0.73 \\
SiO $J$ = 5\nobrl 4 & $[-2.1;+11.4]$ & 1.80 \\
$^{12}$CN $N$ = 2\nobrl 1, $J$ = 5/2\nobrl 3/2\tablefootmark{b} & $[+2.9;+9.4]$ & 1.22 \\
\noalign{\smallskip}
\hline
\end{tabular}
\tablefoot{
	\tablefoottext{a}{$F_1$ = 1\nobrl 0, $F$ = 0\nobrl 1.}
	\tablefoottext{b}{Blend of $F$ = 5/2\nobrl 3/2,  $F$ = 7/2\nobrl 5/2, and $F$ = 3/2\nobrl 1/2.}
}
\end{table}

Figs.~\ref{fig:int1} and \ref{fig:h2co1}--\ref{fig:sio} show molecular line emission in the combined SMA/APEX maps in most of the area covered by both the APEX map and the SMA primary beams. Since primary beam correction is applied, all contours correspond to the same flux level, but not to the same S/N level. Thus, in the channel maps (Figs.~\ref{fig:h2co1}--\ref{fig:sio}) the 0.5~\jybeam\ contours are approximately at a 3$\sigma$ level on the primary beam edges (see Fig.~\ref{fig:smascuba}), but at a 6$\sigma$ level in the beam centres.

The molecular lines seen in the SMA data can roughly be separated into three groups by their distribution in the R~CrA region. The first group consists of all detected H$_2$CO and CH$_3$OH lines. The second group consists of the HC$_3$N, \textit{c}\nobr C$_3$H$_2$, DCN, $^{13}$CN, and $^{12}$CN lines. Finally, the distribution of the SiO emission is not similar to those of the other molecular species.

\subsection{H$_2$CO and CH$_3$OH}

\begin{figure}[!htb]
    \centering
	\includegraphics{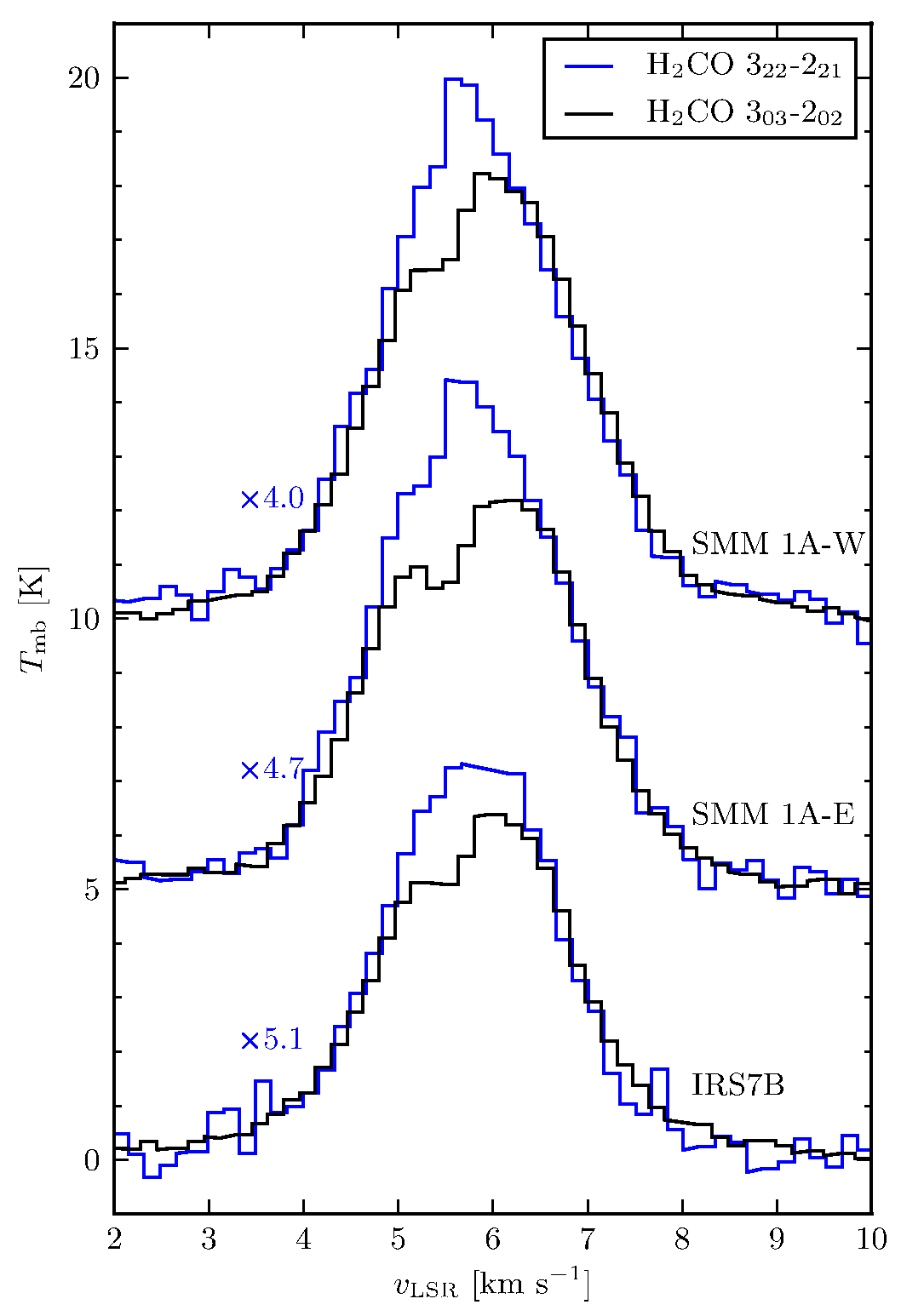}
    \caption{APEX spectra for H$_2$CO 3$_{03}$\nobrl 2$_{02}$ (black) and H$_2$CO 3$_{22}$\nobrl 2$_{21}$ (blue), the latter line is scaled with different factors (4.0, 4.7, and 5.1 from top to bottom) to make the Gaussian tails fit. The vertical shifts between the spectra are 5~K. The H$_2$CO 3$_{21}$\nobrl 2$_{20}$ (not displayed) has a shape similar to the H$_2$CO 3$_{22}$\nobrl 2$_{21}$ line.}
    \label{fig:apex_h2co_compare}
\end{figure}

In the APEX data, the H$_2$CO 3$_{03}$\nobrl 2$_{02}$ line has a non-Gaussian shape not seen in the higher-excited H$_2$CO lines, suggesting either that the emission has multiple components, or that optical depth effects are present (see Fig.~\ref{fig:apex_h2co_compare}).

The H$_2$CO and CH$_3$OH lines show strong emission in two ridges; one south of IRS7B, with one of the peaks in this ridge almost coinciding with the suggested starless core SMM~1A \citep{chen10}; the other between SMM~1C and R~CrA (see Figs.~\ref{fig:int1} and \ref{fig:h2co1}\nobrl \ref{fig:ch3oh}). Both these regions are relatively extended also in the velocity domain ($v_{\mathrm{LSR}} =$~4 to 7~\kms\ for the northern ridge and $v_{\mathrm{LSR}} =$~5 to 8~\kms\ for the southern), but with individual clumps that peak at different velocities in these intervals. Note, however, that significant H$_2$CO 3$_{03}$\nobrl 2$_{02}$ emission is seen all over the field between 4 and 7~\kms. This morphology is different from the sources studied by \citet{jorgensen07}, where H$_2$CO and CH$_3$OH emission mostly is found to be associated with the YSOs and their outflows.

\begin{figure*}[!htb]
    \centering
    $\begin{array}{c@{\hspace{0.0cm}}c@{\hspace{0.0cm}}c}
    \includegraphics{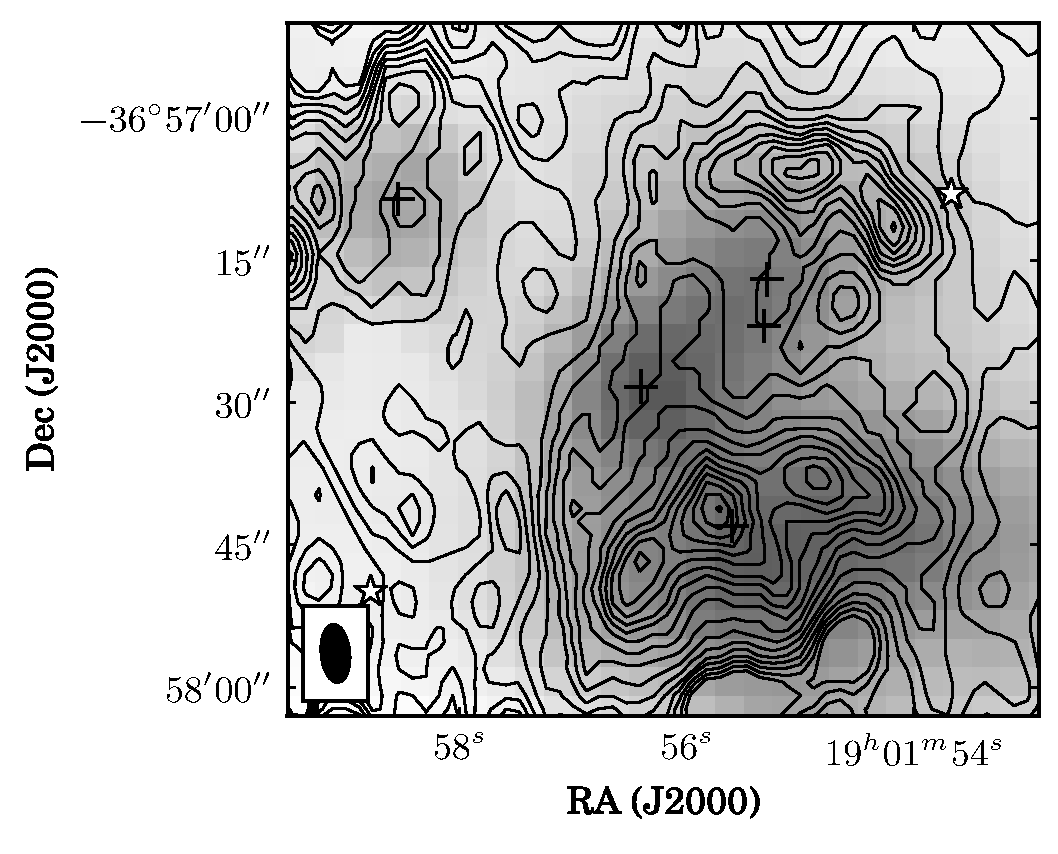} &
    \includegraphics{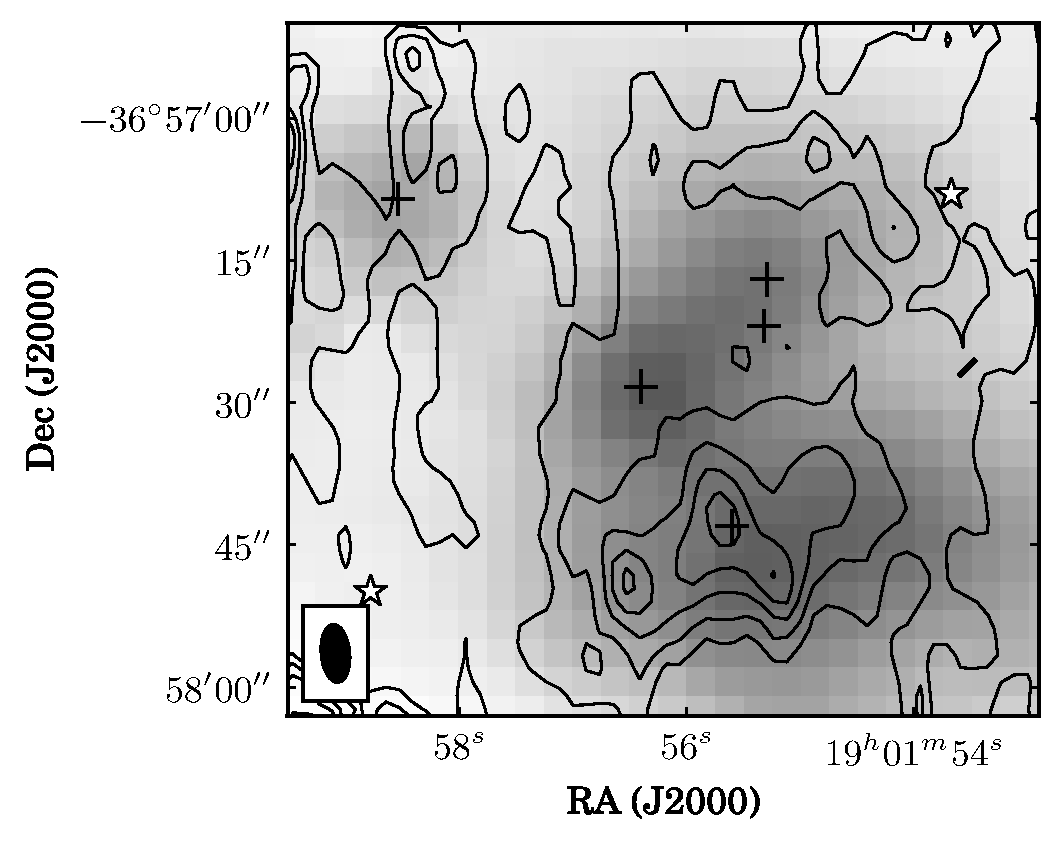} \\
    \end{array}$
    \caption{Integrated intensity of the H$_2$CO 3$_{03}$\nobrl 2$_{02}$ line (left) and H$_2$CO 3$_{22}$\nobrl 2$_{21}$ line (right). Contours every 2~K~\kms\ ($\sim 3\sigma$). Both lines are integrated between 0 and 9~\kms. Greyscale background of both figures is SCUBA 850~\hbox{\textmu}m continuum. Refer to Fig.\ref{fig:herschelsmaspitzer} for a guide to the symbols used for the compact objects.}
    \label{fig:h2coscuba}
\end{figure*}

The H$_2$CO 3$_{03}$\nobrl 2$_{02}$ emission is of comparable strength in the northern and southern ridges, whereas the higher excited H$_2$CO lines and the CH$_3$OH line are stronger in the southern ridge. The difference in the H$_2$CO line intensities suggests a higher temperature in the southern ridge than in the northern or alternatively a higher optical depth of the main transition. The emission peak in the southern ridge does not coincide with the centre of the extended SCUBA dust emission, which lies approximately 10\arcsec\ southwest of the strongest H$_2$CO 3$_{03}$\nobrl 2$_{02}$ peak in the southern ridge. However, the SCUBA dust emission peak coincides well with a secondary peak of the higher excited lines (see Fig.~\ref{fig:h2coscuba}).

For the lower LSR velocities ($-1$ to $+3$~\kms), most H$_2$CO, CH$_3$OH, and SiO emission is seen in the northeastern part of the field covered by the SMA/APEX observations. This low-velocity emission is not connected to the emission around IRS7 -- if that was the case, a much smoother westward translation of the emission with increasing velocity would be seen in Fig.~\ref{fig:h2co1}. Instead, the H$_2$CO emission shows a distinct jump around 3~\kms. This indicates that the northeastern emission originates in another source, outside or on the edge of the field-of-view, perhaps SMM~2. This emission is also fainter than that in the centre of the field.

\subsection{HC$_3$N, \textit{c}\nobr C$_3$H$_2$, DCN, and CN}
The second group of lines (HC$_3$N, \textit{c}\nobr C$_3$H$_2$, DCN, and CN) shows relatively weak emission. Except for CN, they are all present in an EW-extended region between IRS7B and SMM~1A (see Figs.~\ref{fig:int1} and \ref{fig:hc3n}\nobrl \ref{fig:12cn}). \textit{c}\nobr C$_3$H$_2$ and CN are the only molecules in the SMA/APEX data that have their strongest emission associated with the YSOs: \textit{c}\nobr C$_3$H$_2$ peaks at IRS7B, and extends in a northward collimated outflow, whereas $^{12}$CN shows a strong clump just north of IRS7B. Extended HC$_3$N emission is detected south of IRS7A and southwest of IRS7B. The $^{13}$CN emission is rather diffuse and extended, but also of quite low S/N. For the $^{12}$CN lines, no APEX short-spacings are available, so no conclusions about the extended emission can be drawn from this map. However, a strong clump of $^{12}$CN emission is seen peaking just north of IRS7B, and a fainter clump is seen between IRS7A and R~CrA. 

\subsection{SiO}
The SiO emission is present in four distinct regions separated both in position and velocity (see Fig.~\ref{fig:sio}). It appears as a bright point source south of R~CrA, west of SMM~1C, at velocities (9 to 12~\kms) considerably higher than that of R~CrA (5~\kms); in somewhat extended emission around SMM~1A at $v_{\mathrm{LSR}} = 6$~\kms; as a pointlike source west of T~CrA at low velocity ($-1$ to $+2$~\kms); and finally some faint emission east of SMM~2 at a velocity slightly lower than the systemic velocity of R~CrA (0 to 3~\kms). The latter two emission regions are outside the SMA primary beams, and thus only tentative. All SiO peaks coincide with relatively faint H$_2$CO emission, which is in agreement with SMA observations of other prominent embedded protostars showing these molecules to coincide in outflow-affected gas \citep{jorgensen07}.

The most interesting SiO feature is the point source south of R~CrA, since it together with R~CrA falls on a straight line with several Herbig-Haro (HH) objects. \citet{hartigan87} observed several HH objects (HH98, HH98B, HH100, HH97, HH96, and HH101) in a straight line that extends southwestwards from R~CrA, through the SiO peak in the SMA data. They argue that all HH objects are driven by the infrared source IRS1, almost 1\arcmin\ southwest of R~CrA. However, \citet{wang04} argue that, considering the positions of the objects, the outflow might as well be driven by R~CrA. The proper-motion velocities of the objects southwest of IRS1 were measured by \citet{peterson11}, and are indeed directed southwestwards. The proper-motion velocity of HH98B (MHO~2001Q) is also measured, which is directed westwards, and suggested to be driven by SMM~1C (SMA~2), and is thus not related to the other HH objects along the line.

As a summary, the Herbig-Haro objects in the region have several possible driving sources (including IRS1, SMM~1C, and R~CrA), and the SiO peak south of R~CrA could be associated with any of these objects. There are no sources described in \citet{peterson11} that can explain the SiO emission associated with SMM~1A, nor have any HH outflows been observed in its vicinity.

\section{Analysis}
\label{sec:analysis}

With the high resolution interferometric maps from the SMA and the recovery of the more extended structure through the APEX observations, it is possible to investigate the physical conditions in the region around the protostars close to R~CrA. In the following we constrain the physical properties in the region and subsequently (Sect.~\ref{sec:discussion}) the implications concerning the radiation field around the embedded protostars and in particular the influence from R~CrA itself.

\subsection{LTE estimates of physical properties}
\label{sec:lte}
Observations of several rotational transition lines of the same molecule can be used to construct a so-called rotational diagram for a given molecular species from which the column density and rotational temperature of the species can be estimated \citep{goldsmith99}. If one assumes that the molecules are excited under local thermal equilibrium (LTE) conditions and that the lines are optically thin, the rotational temperature is another measure of the kinetic temperature of the species. In the case of H$_2$CO, the three lines at 218~GHz make it possible to calculate first-order approximations of the temperature and column density \citep[see e.g.][]{mangum93,maret04}. However, if the lines are optically thick, the rotational diagram method will usually overestimate the kinetic temperatures, since the lower-excited lines will be more severely affected by the optical thickness than the higher-excited lines.

\begin{figure*}[!htb]
    \centering
    $\begin{array}{c@{\hspace{0.0cm}}c@{\hspace{0.0cm}}c}
    \includegraphics{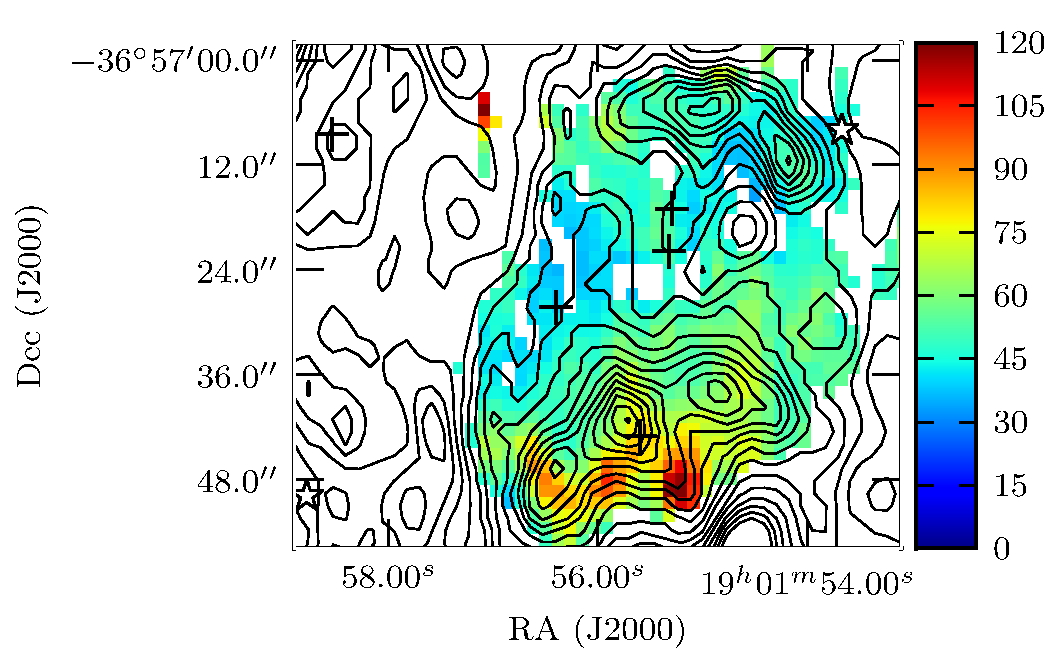} &
    \includegraphics{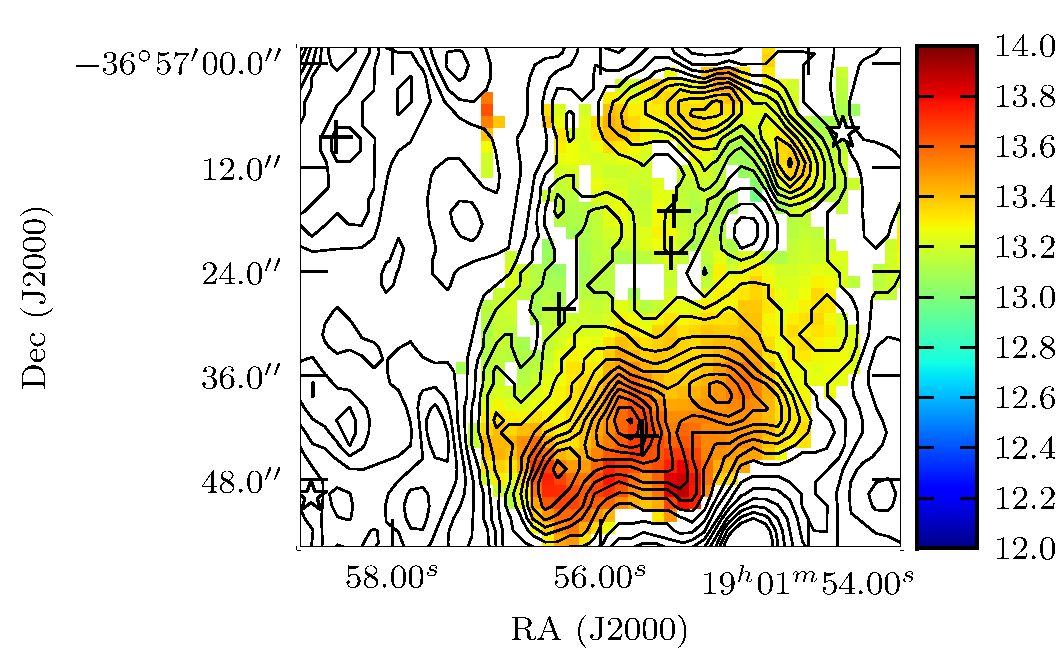} \\
    \end{array}$
    \caption{\textit{Left:} H$_2$CO rotational temperature map of the IRS7 region from SMA/APEX data. Temperatures are in K. \textit{Right:} p-H$_2$CO LTE column density map of the IRS7 region. The colour scale bar shows values in $\log_{10}{N}$, where $N$ is given in cm$^{-2}$. \newline Both maps use the integrated intensities of three para-H$_2$CO lines around 218~GHz, and only data points that are inside the SMA primary beams and where all three lines have $5\sigma$ detections or stronger are considered. All other points are represented by white pixels. The temperatures and column densities are calculated with rotational diagrams for line intensities integrated between 0 and 9~\kms. The contours show the integrated intensity of the H$_2$CO 3$_{03}$\nobrl 2$_{02}$ line in the same velocity interval, with contours every 2~K~\kms\ ($\sim 3\sigma$ at the edge of the primary beam). Refer to Fig.\ref{fig:herschelsmaspitzer} for a guide to the symbols used for the compact objects.}
     \label{fig:h2co_tc}
\end{figure*}

The temperature and column density of the H$_2$CO in the region covered by the SMA and APEX data is estimated by the use of these three spectral lines in all parts of the SMA primary beams where the integrated intensities between 0 and 9~\kms\ of all three spectral lines have a signal-to-noise ratio of at least 5. In Fig.~\ref{fig:h2co_tc}, the rotational temperatures and column densities estimated from the SMA/APEX maps of the H$_2$CO 3$_{03}$\nobrl 2$_{02}$, 3$_{21}$\nobrl 2$_{20}$, and 3$_{22}$\nobrl 2$_{21}$ lines are shown. Note that the lines in this study are para-H$_2$CO lines, and to find the total H$_2$CO column densities, the ortho/para ratio must be accounted for. In this paper, the p-H$_2$CO column density will be stated. Assuming an ortho/para ratio of $1.6$ \citep{dickens99,jorgensen05}, the total H$_2$CO column density can be estimated by multiplying the p-H$_2$CO column density by $2.6$. 

\begin{figure}[!htb]
    \centering
    \includegraphics{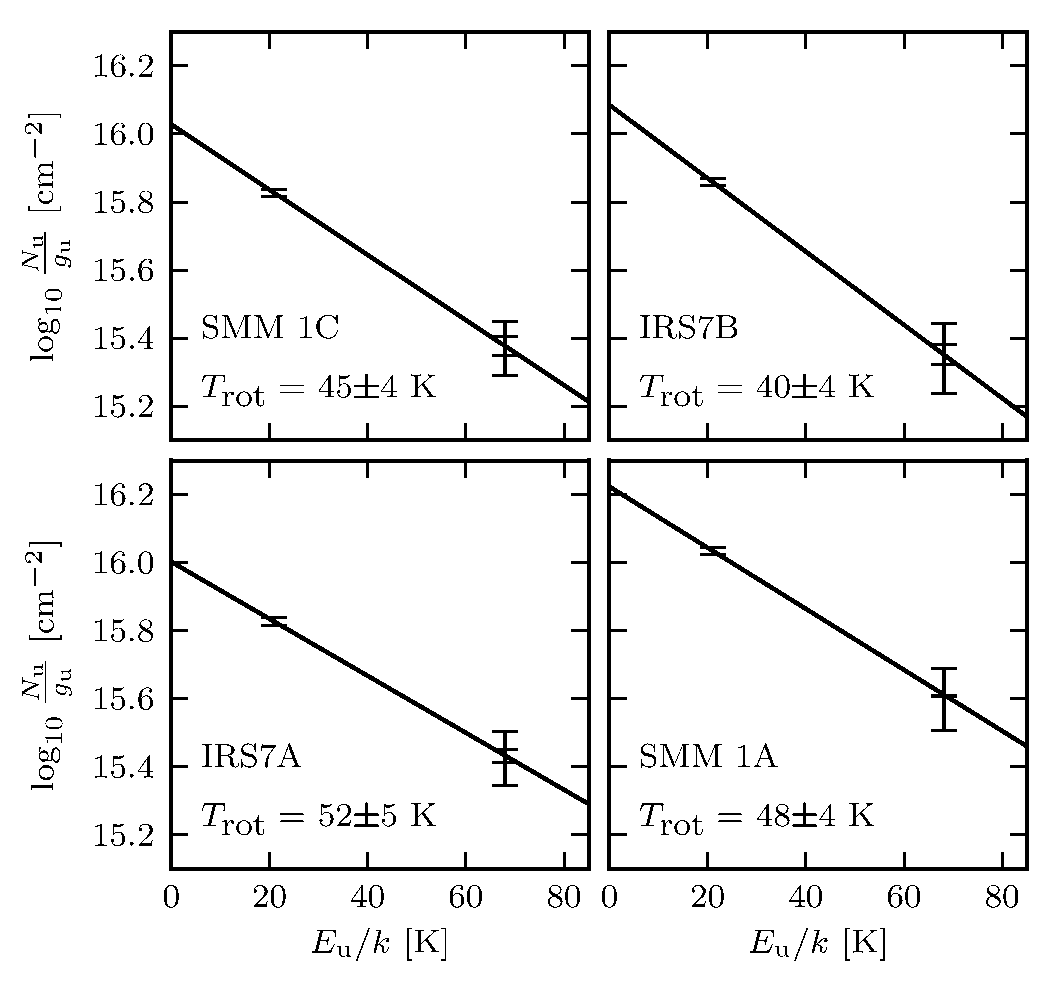}\\
    \caption{H$_2$CO rotation diagrams for the YSOs in the SMA/APEX field (SMM~1C, IRS7B, IRS7A and SMM~1A).}
     \label{fig:rotdiag_yso}
\end{figure}

The maps show a wide spread in rotational temperatures throughout the region; from 30~K in the colder regions (mostly associated with low H$_2$CO intensities, and thus low S/N levels), up to 120~K in a few peaks close to SMM~1A. From the SMA/APEX map of the IRS7 cloud (Fig.~\ref{fig:h2co_tc}), it can be found that the highest rotational temperatures are not associated with the YSOs, but with the southern ridge around SMM~1A, and also with the northern ridge close to R~CrA. In some points in the southern ridge, a rotational temperature of $120\pm20~$K is found. Using the APEX pointing centred at IRS7B, a single-dish rotational temperature of 62~K was calculated in a 28.6\arcsec\ beam around IRS7B. However, the highest temperatures are likely optical depth effects, and the physical temperatures should rarely exceed 70~K (see Sect.~\ref{sec:opticaldepth}). The estimated column density varies between $10^{13}$ and $10^{14}$~cm$^{-2}$ in the region, and is fairly well-correlated with the rotational temperature. As can be seen in Fig.~\ref{fig:h2co_tc}, the rotational temperature is not peaking at the same place as the H$_2$CO 3$_{03}$\nobrl 2$_{02}$ (main line) emission. The rotation temperatures at the positions of the YSOs are relatively moderate, around 40--50~K (see Fig.~\ref{fig:rotdiag_yso}).

Only one of the three high-temperature H$_2$CO peaks around SMM~1A coincides with one of the three condensations within SMM~1A suggested by \citet{chen10}: SMM~1A\nobr a, the weakest 1.3~mm continuum source within SMM~1A, coincides with the peak with the second highest rotational temperature. The much more luminous SMM~1A\nobr c is found in a point with relatively low H$_2$CO temperature ($T\approx 65$~K). The northern ridge of extended H$_2$CO emission has a more even and somewhat lower rotational temperature ($T\approx$~40--50~K) than the southern ridge ($T\approx$~50--120~K).

These temperature and column density estimates suffer from four main sources of errors: The statistical error (noise); the calibration error of the telescopes; the primary beam error, which is notable far from the primary beam centres; and the errors arising from the model assumption of LTE conditions. The statistical uncertainty, including the primary beam error, is estimated from the RMS of the noise in the combined SMA and APEX data. The systematic calibration error is estimated to $20\%$ of the line flux (note that both SMA and APEX will introduce systematic errors, which both should be in the order of $20\%$, but since the line flux is dominated by the APEX data (see Fig.~\ref{fig:short_compare2}), it can be assumed that the total calibration error is almost equal to the contribution from APEX). However, this error does only apply to the column density estimates, since the rotational temperatures are proportional to the logarithm of the line ratios and thus only depend on the relative errors in the line flux estimates due to noise. Typically, the rotational temperatures are therefore determined to about $10\%$ accuracy given the model assumptions.

\citet{maret04} studied the H$_2$CO abundances in a sample of Class~0 protostars using the IRAM 30~m telescope and the JCMT. Several p-H$_2$CO and o-H$_2$CO spectral lines between 140~GHz and 365~GHz were observed in eight sources, and the highest H$_2$CO rotational temperatures were found in three hot corino sources in NGC~1333 (25--40~K), whereas the rotational temperatures in the other sources were in the order of 10--20~K. However, if only using the 218~GHz lines, much higher rotational temperatures (up to several 100~K) are reached in the hot corino sources, but with large uncertainties. Furthermore, \citet{jorgensen05,jorgensen07} demonstrated that the single-dish H$_2$CO observations are difficult to interpret for the specific sources -- in particular, that the high rotational temperatures are a likely result of extended outflow emission picked up by the single-dish observations.

\subsubsection{Optical depth}
\label{sec:opticaldepth}

\begin{figure}[!htb]
    \centering
	\includegraphics{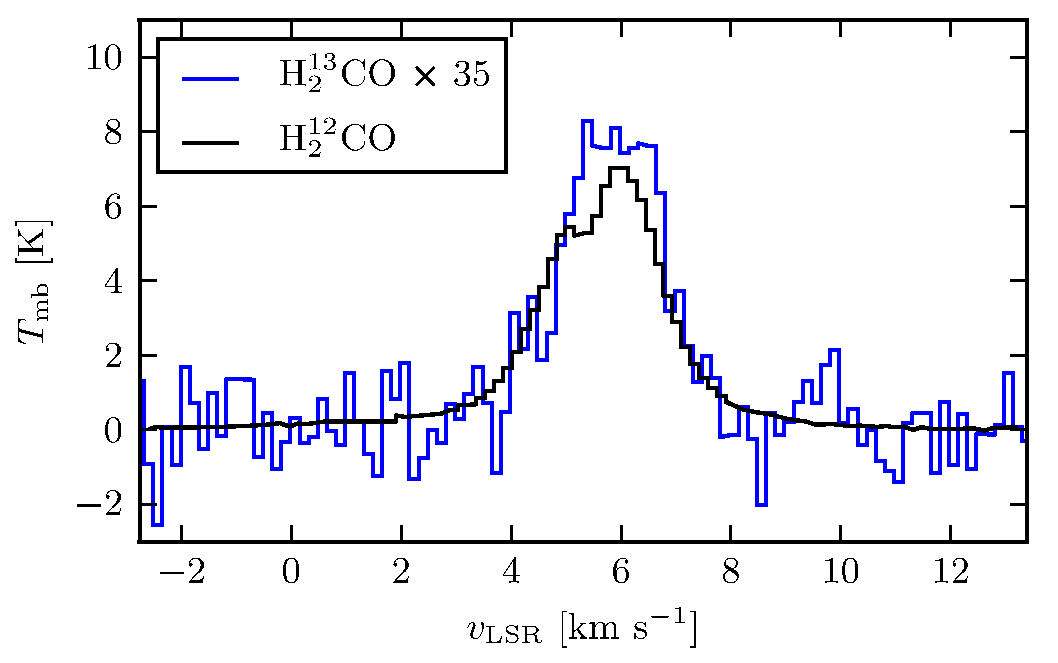}
    \caption{H$_2^{12}$CO 3$_{12}$\nobrl 2$_{11}$ (black) and H$_2^{13}$CO 3$_{12}$\nobrl 2$_{11}$ (blue, scaled by the factor 35) in IRS7B. Data from APEX line survey of IRS7B (Lindberg et~al., in prep.).}\label{fig:h213co}
\end{figure}

To investigate whether the assumption of optically thin H$_2$CO lines is valid, the intensities of the H$_2^{12}$CO 3$_{12}$\nobrl 2$_{11}$ line at 225.6978~GHz and the H$_2^{13}$CO 3$_{12}$\nobrl 2$_{11}$ line at 219.9085~GHz (both from a large APEX survey; Lindberg et~al. in~prep.) were compared. The ratio between the integrated intensities of the two lines was calculated to be 35 (Fig.~\ref{fig:h213co}), which should be compared to the typical ISM $^{12}$C/$^{13}$C ratio in the local ISM of $77\pm7$ \citep{wilson94}. The ratio measured in IRS7B is thus a factor of 2.2 lower, giving an optical depth of the o\nobr H$_2^{12}$CO 3$_{12}$\nobrl 2$_{11}$ line of $\tau \approx 0.8$. Using equations for the total column density of a molecule \citep[see e.g.][]{goldsmith99}, and assuming an ortho/para ratio of 1.6 \citep{dickens99,jorgensen05}, and $T_{\mathrm{ex}} = 62$~K (the rotational temperature observed in the APEX beam), we find that this corresponds to $\tau \approx$~0.69 for the p\nobr H$_2$CO 3$_{03}$\nobrl 2$_{02}$, and $\tau \approx$~0.18 for the p\nobr H$_2$CO 3$_{21}$\nobrl 2$_{20}$ and p\nobr H$_2$CO 3$_{22}$\nobrl 2$_{21}$ lines. Note, however, that these optical depths are averages over the APEX 28.6\arcsec\ beam, and that locally elevated optical depths are likely.

Since the optical depths of the three spectral lines used for the rotational temperature calculation are different, the rotational temperature will be overestimated. When compensating for optical depth effects using the values estimated above and the optical depth correction factor given by \citet{goldsmith99}, the rotational temperature in the APEX beam drops from 62~K to 47~K. Since the distribution of the optical depth across the SMA/APEX map is unknown, non-LTE methods must be used to acquire a more accurate temperature estimate, but applying the large-scale optical depth on the calculations of the rotational temperature peaks in the southern ridge makes these drop from 100--120~K to 65--75~K.

\subsection{Non-LTE analysis}
\label{sec:radex}

For further analysis of the emission, we utilised the RADEX escape probability code \citep{vandertak07} to solve the excitation of H$_2$CO under non-LTE conditions assuming a uniform spherical molecular cloud. Besides allowing for a more detailed comparison of the importance of the density and temperature, this method also provides a path to take optical depth effects into account. Collision rates come from \citet{green91}, and were retrieved from the LAMDA database \citep{lamda}.

In the absence of a central heating source, the temperature of a molecular cloud can hardly exceed 10--15~K assuming no external heat contribution, and typically the density would be $\sim10^4$--$10^5$~cm$^{-3}$ \citep{bergin07}. Furthermore, the low-mass YSOs should not be able to heat the gas on scales observable in the SMA/APEX map. Consequently, assuming external heating, a fairly even temperature distribution is expected across each dense ridge. Thus, it is likely that the wide spread of temperatures found in the rotational temperature map (Fig.~\ref{fig:h2co_tc}) is caused exclusively by column density effects, and that each of the two ridges have uniform temperatures. To exclude all emission outside these ridges, only pixels where all three H$_2$CO lines are detected to at least an $8 \sigma$ level are considered in the RADEX calculations.

In the ridges, the H$_2$CO 3$_{21}$\nobrl 2$_{20}$/H$_2$CO 3$_{22}$\nobrl 2$_{21}$ ratio has an average of 0.98, and the ratio lies between 0.8 and 1.2 in more than $90\%$ of the pixels in the two ridges. The average optical depth of the H$_2$CO 3$_{03}$\nobrl 2$_{02}$ line in an APEX beam around IRS7B is 0.7 (see Sect.~\ref{sec:opticaldepth}), but might be several times higher in the peaks. These parameters are found to be relatively temperature-insensitive -- at 30--90~K they require $n \ga 10^5$~cm$^{-3}$ and $N \sim 10^{13}$--$10^{15}$~cm$^{-2}$. Where the H$_2$CO 3$_{21}$\nobrl 2$_{20}$/H$_2$CO 3$_{22}$\nobrl 2$_{21}$ ratio is greater than 1, $n \ga 10^6$~cm$^{-3}$. This suggests that the H$_2$ density in both ridges must be relatively high compared to what one would expect in a prestellar cloud.

RADEX calculations are run on each of the individual pixels in the ridges to investigate which uniform temperature fits the data best. This non-LTE model is tested with uniform H$_2$ densities of $10^5$, $10^6$, $10^7$, and $10^8$~cm$^{-3}$, and the column density is set to be a free parameter across the cloud. The two ridges are treated separately, and the most likely temperature of each ridge is established by calculating the $\chi^2$ of the intensities modelled by RADEX compared to the observed intensities, and then minimising the median of these $\chi^2$ values for all pixels in each ridge. The temperatures where the optical depth of any of the lines is negative are excluded (a negative $\tau$ corresponds to masing, which is very unlikely to occur on these large scales, which removes some high-temperature solutions). Temperatures where the optical depth of the H$_2$CO 3$_{03}$\nobrl 2$_{02}$ line is greater than 10 are also discarded (this is more than an order of magnitude larger than the average value estimated in Sect.~\ref{sec:opticaldepth}, which excludes some low-temperature solutions).

\begin{figure}[!htb]
    \centering
	\includegraphics{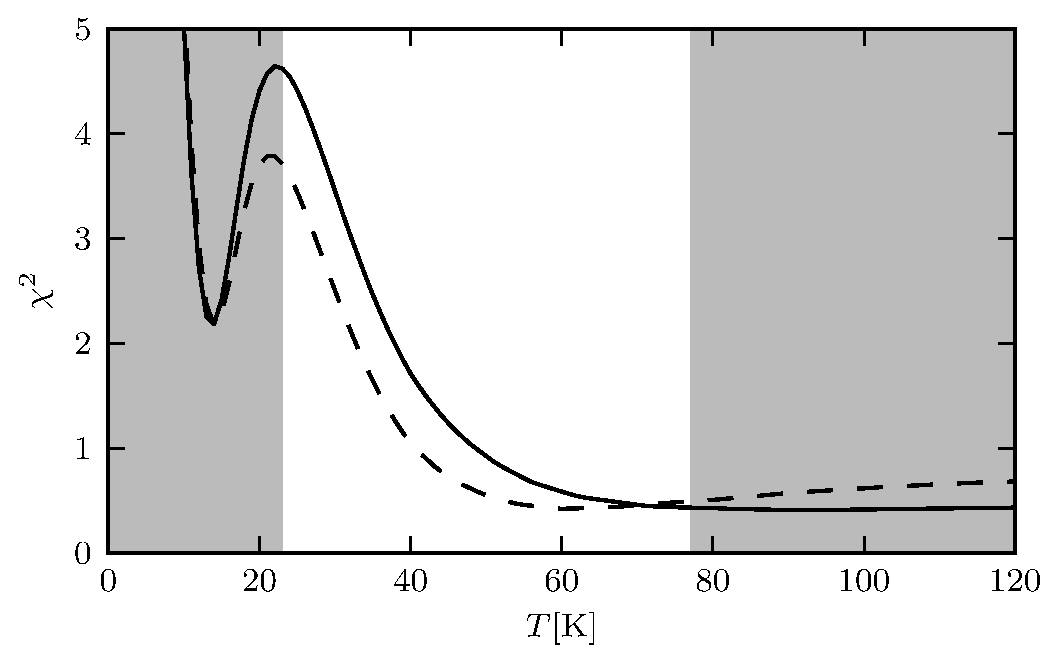}
    \caption{$\chi^2$ fits to RADEX data and the observed H$_2$CO lines to a constant temperature with variable column density (see Fig.~\ref{fig:radex_ridges_coldens}) and an H$_2$ density of $10^6$~cm$^{-3}$ for the northern ridge (dashed) and the southern ridge (solid). The greyed areas in the plot correspond to solutions with $\tau > 10$ (left) and $\tau < 0$ (right).}
    \label{fig:upperlower_temp}
\end{figure}

\begin{figure*}[!htb]
    \centering    
    $\begin{array}{c@{\hspace{0.0cm}}c@{\hspace{0.0cm}}c}
    \includegraphics{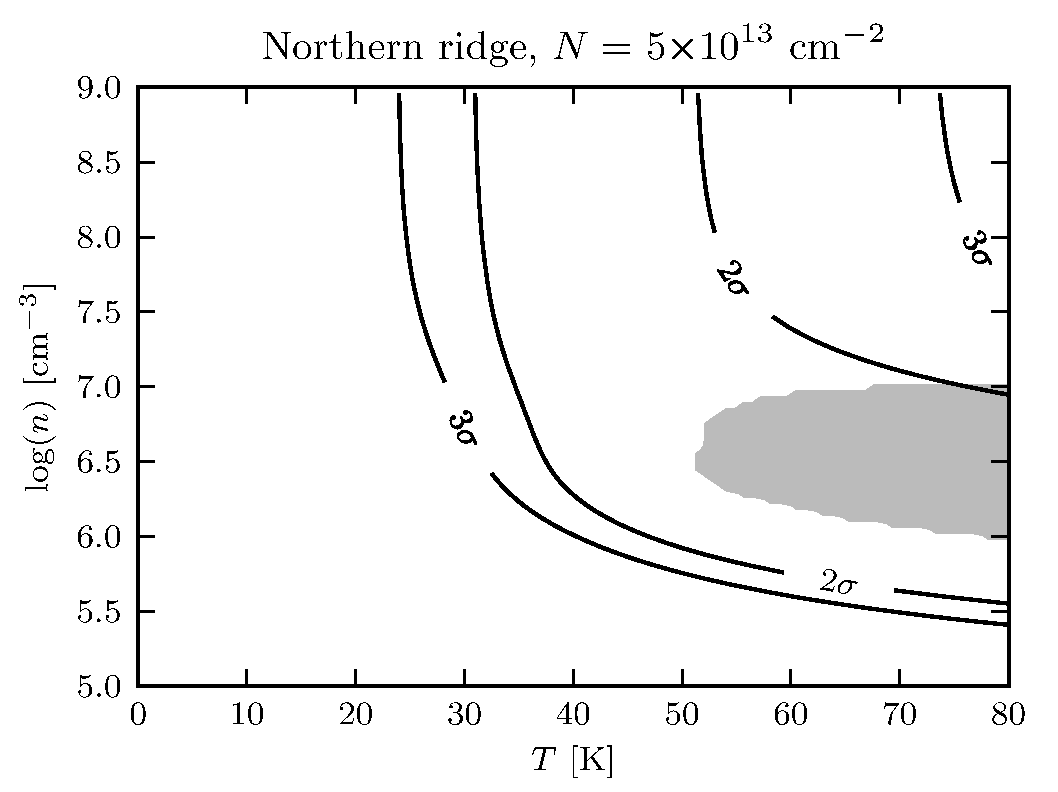} &
    \includegraphics{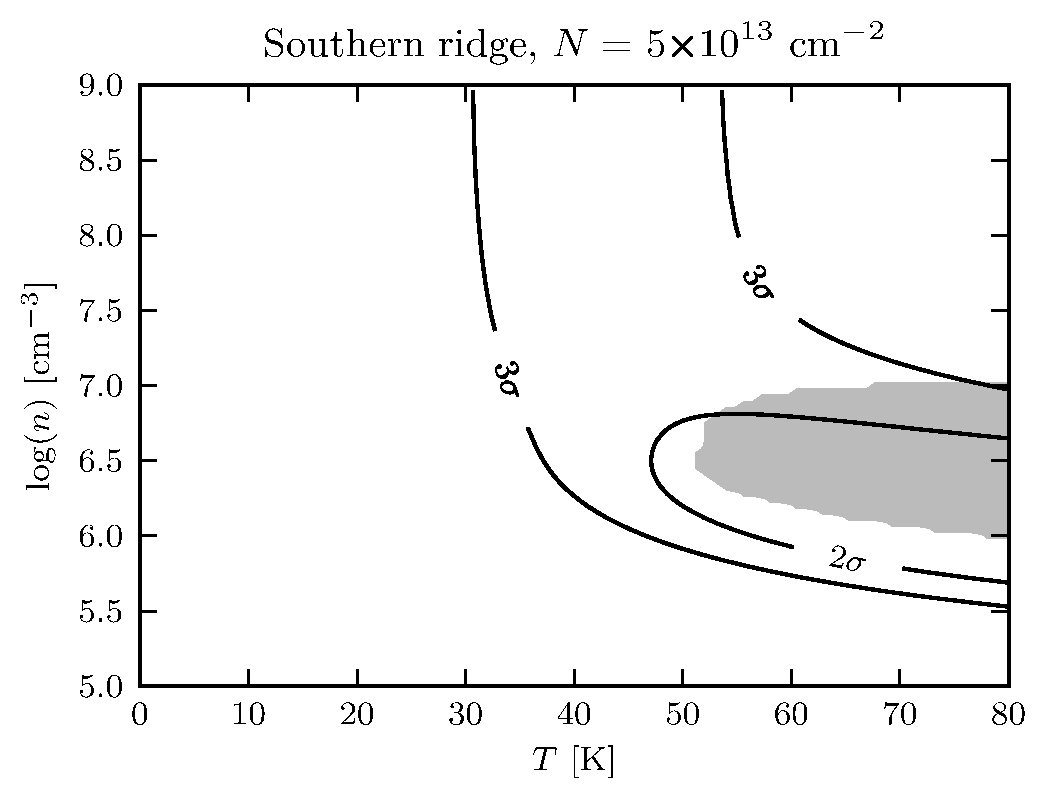} \\
    \end{array}$
    \caption{$\chi^2$-confidence contour plots for temperature and number density determined from the observed H$_2$CO lines and RADEX models. \textit{Left:} Fit to the average intensity in the northern ridge, $N=5\times10^{13}$~cm$^{-2}$. \textit{Right:} Fit to the average intensity in the southern ridge, $N=5\times10^{13}$~cm$^{-2}$. The greyed areas represent solutions with masing H$_2$CO emission.}
    \label{fig:chi2_ridges}
\end{figure*}

\begin{figure}[!htb]
    \centering
	\includegraphics{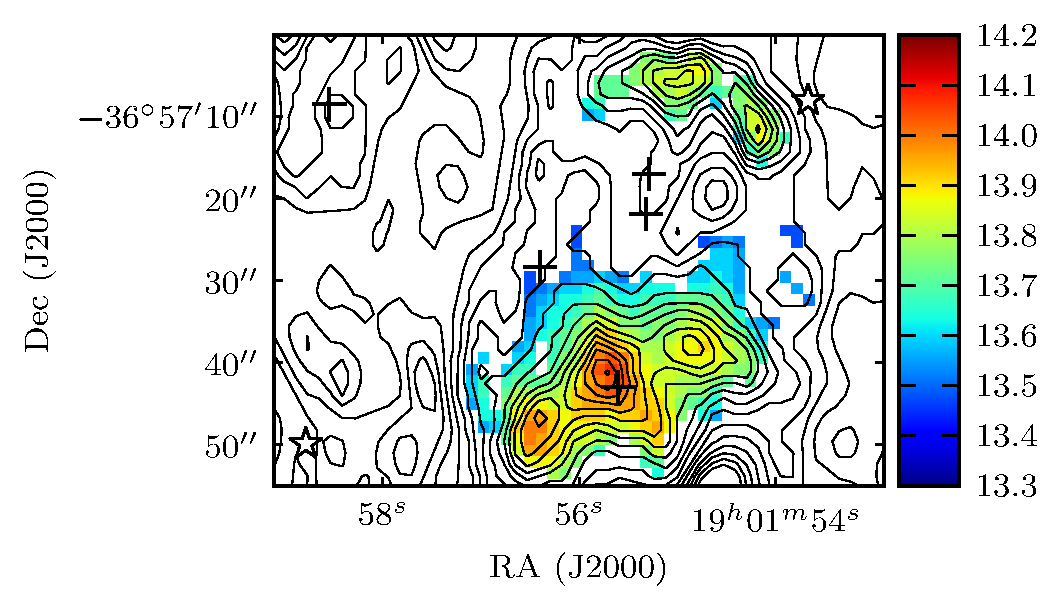}
    \caption{p-H$_2$CO column densities for the least-$\chi^2$ fits to RADEX data and the observed H$_2$CO lines to a constant temperature in each ridge (60~K in northern ridge and 76~K in southern ridge) and an H$_2$ density of $10^6$~cm$^{-3}$. The colour bar shows values in $\log_{10}{N}$, where $N$ is given in cm$^{-2}$. Refer to Fig.\ref{fig:herschelsmaspitzer} for a guide to the symbols used for the compact objects.}
    \label{fig:radex_ridges_coldens}
\end{figure}

From the RADEX calculations, it is found that the density must be higher than $10^5$~cm$^{-3}$, since densities lower than that require a very high optical thickness ($\tau>10$) to produce the observed intensities in several positions at reasonable temperatures ($T<90$~K). The optical depths found at $10^6$~cm$^{-3}$ are more likely, but for the highest temperatures ($T\ga70$~K), one or several of the lines would be masing. The lowest-$\chi^2$ temperatures (excluding $\tau < 0$ and $\tau > 10$) for each of the fits can be found in Table~\ref{tab:radextemps}, where it can be seen that the best fits give temperatures of at least 50~K in the southern ridge in any scenario (see also Fig.~\ref{fig:upperlower_temp}). From Fig.~\ref{fig:chi2_ridges}, it can also be found that given the most likely column densities estimated from the optical depth (see Sect.~\ref{sec:opticaldepth}), the temperature is higher than 25~K in the northern ridge and higher than 30~K in the southern ridge with $3\sigma$ certainty. Consequently, both the temperatures and densities in the region are likely considerably higher than the 10--15~K and $10^4$--$10^5$~cm$^{-3}$ expected in pre-stellar cores \citep{bergin07,wardthompson07}. The best-fit column densities are shown in Fig.~\ref{fig:radex_ridges_coldens}, and lie around $10^{13}$--$10^{14}$~cm$^{-2}$ in both ridges (slightly higher in the southern ridge than in the northern).

\begin{table}[!htb]
\centering
\caption[]{Temperatures calculated by the non-LTE code RADEX. The given temperature corresponds to the lowest-$\chi^2$ fit, and the limits correspond to values where $\chi^2 < 2.3$. The column densities are shown in Fig.~\ref{fig:radex_ridges_coldens}.}
\label{tab:radextemps}
\begin{tabular}{c c c}
\noalign{\smallskip}
\hline
\hline
\noalign{\smallskip}
$n$ & $T$ in N ridge & $T$ in S ridge \\
$\left[\mathrm{cm}^{-3}\right]$ & [K] & [K] \\
\noalign{\smallskip}
\hline
\noalign{\smallskip}
$10^5$ & --\tablefootmark{a} & --\tablefootmark{a} \\
$10^6$ & $60^{+17}_{-29}$\tablefootmark{b} & $76^{+1}_{-31}$\tablefootmark{b} \\
$10^7$ & $65^{+2}_{-26}$\tablefootmark{b} & $67^{+0}_{-19}$\tablefootmark{b} \\
$10^8$ & $47^{+45}_{-16}$ & $61^{+77}_{-23}$ \\
\noalign{\smallskip}
\hline
\end{tabular}
\tablefoot{
     	\tablefoottext{a}{$\tau > 10$ for the H$_2$CO 3$_{03}$\nobrl 2$_{02}$ line for all $T$ where median $\chi^2 < 2.3$.} \\
     	\tablefoottext{b}{The higher temperature limit corresponds to the limit where at least one of the lines starts to mase ($\tau < 0$).} \\     	
}
\end{table}

In all the fits, a dip in the $\chi^2$ fit around 15~K is found. This is not the minimum-$\chi^2$, but still within the $1\sigma$ limit. However, those low-temperature fits always require very high optical depths ($\tau > 100$ for $n\leq10^6$~cm$^{-3}$, $\tau > 25$ for $n\geq10^7$~cm$^{-3}$), which is one to two orders of magnitude higher than the optical depth averaged over an APEX beam centred at IRS7B (see Sect.~\ref{sec:opticaldepth}). If we assume an H$_2$CO/H$_2$ ratio in the order of $10^{-9}$ \citep{jorgensen05}, and that the source is spherical and has a size in the order of 1\arcsec, the observed line strengths would require an H$_2$ density in the spherical cloud in the order of $10^9$--$10^{10}$ cm$^{-3}$ assuming the low-temperature solution. The high-temperature solution is thus much more likely.

\section{Discussion}
\label{sec:discussion}

\subsection{Internal vs. external heating}
\label{sec:transphere}

It was not possible to fit the observed line intensities with temperatures lower than 35~K in either the LTE models or the non-LTE models of the physical properties of the cloud. More likely, the temperatures in both ridges lie around 50--60~K, and perhaps as high as 90--100~K in a few clumps south of SMM~1A. SMM~1C lies on the southern border of the northern ridge, but IRS7A and IRS7B do not belong to any of the ridges. In the regions between the ridges, the column densities are approximately an order of magnitude lower than in the ridges. The temperatures around the YSOs are likely also somewhat lower (30--50~K) than in the ridges, but these estimates are not as reliable, since the S/N levels are lower in these regions.

According to dust radiative transfer calculations, the temperature of the envelope heated only by an internal 20~$L_{\sun}$ Class 0/I YSO will at most reach at most 30~K on the scales of 1\,000~AU \citep[see, e.g.,][and references therein]{jorgensen06}. For protostars with even lower luminosities, such as the YSOs in IRS7, the internal heating should be of even lesser importance, and on the scales of the resolution of the SMA data (400--800~AU) temperatures in the order of 50--60~K as a result of internal heating are unrealistic. However, if heating from an external UV field is applied, the models discussed below show that it is possible to reach the measured temperatures.

Not only does the region south of IRS7 show high H$_2$CO temperatures, the 850~\hbox{\textmu}m dust emission also has a strong peak southwest of the strongest H$_2$CO peak (see Fig.~\ref{fig:h2coscuba}). This dust emission peak coincides well with the temperature peak ($T\approx 100$~K) in the rotational temperature map (Fig.~\ref{fig:h2co_tc}). Thus, the higher rotational temperature measured in the peaks could also be an optical depth effect caused by the high column density.

The angular separation between R~CrA and SMM~1A is approximately 50\arcsec, which corresponds to a projected separation of 6\,500~AU assuming the distance of 130~pc to the cloud \citep{neuhauser08}. If the luminosity of R~CrA is assumed to be $\sim200~L_{\sun}$ \citep{bibo92}, its surface temperature should be in the order of $10\,000$~K. Assuming a black body with this temperature, slightly less than $50\%$ of its luminosity ($\sim90~L_{\sun}$) would fall in the UV range between 300~nm and 13.6~eV. Such a star would increase the interstellar radiation field (ISRF) by a factor of $\chi_{\mathrm{ISRF}} \sim3\,000$ at the distance 6\,500~AU, assuming an interstellar radiation field of $8\times10^{-4}$~erg~cm$^{-2}$~s$^{-1}$ (twice the Habing flux). For a slightly lower luminosity for R~CrA \citep[$100~L_{\sun}$ is the lower limit given by ][]{bibo92}, and the distance being somewhat higher (if R~CrA and SMM~1A are not in the same plane of sky), $\chi_{\mathrm{ISRF}} \sim750$. Such high UV fluxes would create a Photon Dominated Region (PDR), which would give rise to high CN abundances, in agreement with \citet{watanabe12}.

To estimate the relative importance between internal and external heating, we constructed a 1-dimensional model for the envelope around IRS7B using the dust radiative transfer code \textit{Transphere} \citep{transphere}. We assumed a power-law density profile for the envelope with \mbox{$n \propto R^{-1.5}$} and then varied the density profile normalisation, envelope size, central source luminosity, and interstellar radiation field. We adopted the dust opacities from \citet{ossenkopf94} corresponding to coagulated dust grains with thin ice mantles (OH5). \textit{Transphere} calculates the self-consistent temperature in the spherical envelope with these parameters. For comparison to the observations we subsequently used the SKY ray-tracer from the RATRAN radiative transfer code \citep{ratran} to calculate continuum maps at different wavelengths. To allow for comparison with SCUBA and \textit{Herschel} SED fluxes, the envelope dust emission at the corresponding telescope primary beam edges was subtracted from the peak emission. This difference was then compared to measured SCUBA \citep{nutter05} and \textit{Herschel} (Lindberg et al., in prep.) continuum fluxes.

\begin{figure}[!htb]
    \centering
	\includegraphics{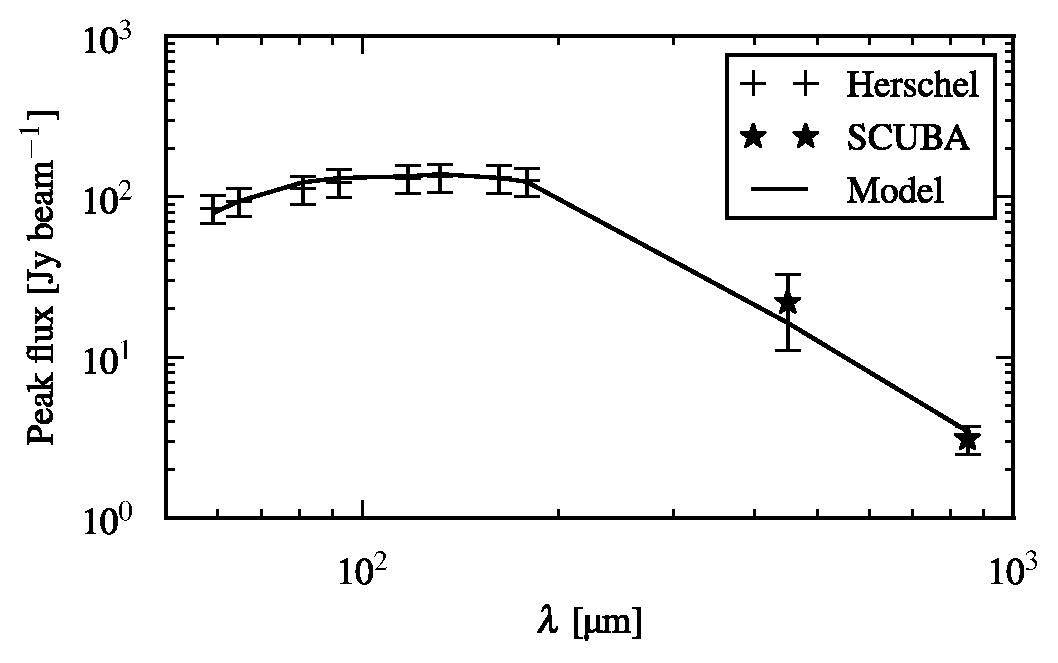}
    \caption{The SED calculated from the envelope with a $4~L_{\sun}$ central source and an interstellar radiation field with $\chi_{\mathrm{ISRF}} = 750$ by the use of RATRAN (the solid line in Fig.~\ref{fig:temp_profile}) compared with the real data points from \textit{Herschel} and SCUBA continuum measurements.}
    \label{fig:sed_model}
\end{figure}

The SED produced by the best fit (with both internal and external heating) is compared to the SCUBA and \textit{Herschel} continuum fluxes in Fig.~\ref{fig:sed_model}. The eight wavelengths at which the \textit{Herschel} fluxes were estimated were chosen to cover each of the four PACS bands with two data points equidistant from the band edges and each other. The systematic continuum data errors are estimated to $20\%$ of the fluxes for the \textit{Herschel} data (Green et al., in prep.), and $50\%$ and $20\%$ for the 450~\hbox{\textmu}m and 850~\hbox{\textmu}m SCUBA data, respectively \citep{difrancesco08}. The best fit was produced for a central source of $4~L_{\sun}$, with an envelope between $R_{\mathrm{in}} < 200$~AU and $R_{\mathrm{out}} = 10\,000$~AU with \mbox{$n \sim R^{-1.5}$}, $n_0 = 1.3 \times 10^6$~cm$^{-3}$ at the reference radius $R_0 = 1\,000$~AU (corresponding to an envelope mass of $2.2~M_{\sun}$), and an interstellar radiation field with $\chi_{\mathrm{ISRF}} = 750$.

\begin{figure}[!htb]
    \centering
	\includegraphics{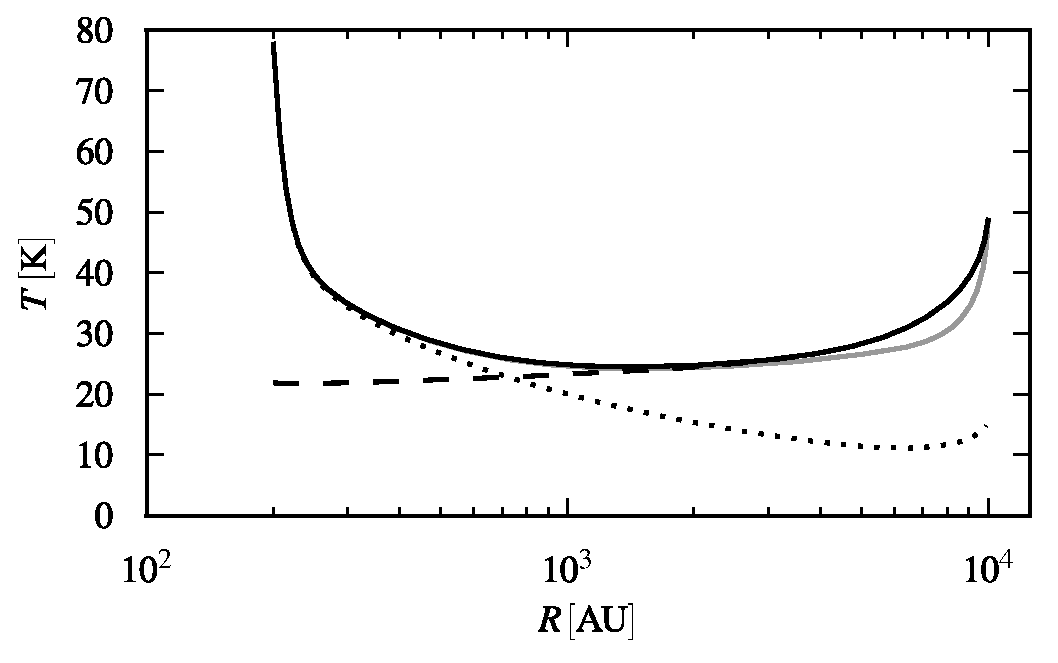}
    \caption{Temperature as a function of radius for the four different models produced with \textit{Transphere}. 
    Black solid line: With a central source of $4~L_{\sun}$ and an interstellar radiation field with $\chi_{\mathrm{ISRF}} = 750$. Dotted line: With a central source of $4~L_{\sun}$ and a normal interstellar radiation field ($\chi_{\mathrm{ISRF}} = 1$). Dashed line: Without a central source, but with an interstellar radiation field with $\chi_{\mathrm{ISRF}} = 750$. Grey solid line: As the black solid line, but with an ambient density of $10^5$~cm$^{-3}$. }
    \label{fig:temp_profile}
\end{figure}

The calculated temperature profile is shown in Fig.~\ref{fig:temp_profile}, where $T\approx25$~K for $R\approx2\,500$~AU, the distance between IRS7B and SMM~1A. Also shown is the model without the external radiation field: as argued above, in this model the temperature drops below 20~K at radii larger than about 700~AU, thus not reproducing the large-scale temperatures observed in SMA/APEX data. In contrast, in the model with the enhanced external radiation, the temperature never drops below 25~K, and increases reaching 50~K in the outermost regions of the envelope. With an enhanced external radiation field ($\chi_{\mathrm{ISRF}} = 750$) and no internal source of heating, the temperature stays above 20~K in the whole envelope.

Naturally, this 1-dimensional model is a simplified description of the physical properties in the region. For example, it is unlikely that the density profile continues to follow the power-law at large radii. Since we see H$_2$CO emission in the whole field, the density should at least exceed the effective density of the H$_2$CO transitions, i.e. at least be in the order of $10^5$~cm$^{-3}$ \citep{evans99}. Adding an ambient density of $10^5$~cm$^{-3}$ to the density profile only marginally changes the temperature profile (Fig.~\ref{fig:temp_profile}).

The 1-dimensional model is also a simplification since R~CrA in reality would cause a radiation field that decreases from northwest to southeast, instead of acting constantly from all directions. The fact that the northern ridge has a lower measured temperature than the southern ridge, despite having a smaller angular separation with R~CrA, is most likely a geometric effect. Either it is caused by extinction and local density distribution differences, or the ridges simply are positioned at different distances from us, making the actual distance between the northern ridge and R~CrA larger than that between the southern ridge and R~CrA. The difference in LSR velocities between the ridges do suggest that they are situated in different planes of sky. Still, this simplified model is useful for a first interpretation of the data at hand. It shows that the relatively high temperatures in the ridges cannot be explained by the internal heating from the protostars, but that they instead can be explained by R~CrA acting as an external heating source upon the envelope.

\subsection{Implications}

As demonstrated based on the radiative transfer models (Sect.~\ref{sec:transphere}), the temperature in the protostellar envelope of IRS7B is currently above 20~K due to the irradiation by R CrA. This may have important implications for the chemistry of the envelope and thus potentially also the composition of the the emerging circumstellar disc.

The gas-phase chemistry of the protostellar envelope would be dominated by radicals such as CN and C$_2$H prominent in PDRs or regions of protostellar envelopes affected by strong UV irradiation \citep[e.g.,][]{jansen95,sternberg95,fuente95,jorgensen04_l483}. In fact, in the line survey of IRS7B by \cite{watanabe12}, an over-abundance of these species is seen. 

The strong UV irradiation may, however, also affect the grain surface chemistry, potentially leading to the formation of more complex organic species. If R~CrA has been as luminous throughout the evolution of IRS7B (from the time of its dense prestellar core stage) this would have hindered significant freeze-out of molecules such as CO, CH$_4$, and N$_2$, being some of the starting points for grain-surface reactions leading to zeroth generation species such as CH$_3$OH \citep{herbst09}. However, if R~CrA itself has undergone significant evolution in terms of luminosity it may just as well have stimulated the formation of complex organic molecules through the photochemistry in CH$_3$OH-rich ices \citep{oberg09}. With the present observational data, it is difficult to distinguish these scenarios: the low internal luminosity of IRS7B means that the regions where complex organic species will return to the gas-phase are small ($\la 100$~AU) and severely diluted in typical single-dish beams. Future high angular resolution observations with ALMA will, however, reveal emission of complex organic molecules if present.

\section{Conclusions}

We have presented high angular resolution observations of young stars in the region of R~CrA from the Submillimeter Array augmented by short-spacing maps from the APEX single-dish telescope. Together the observations reveal the physical and chemical structure of the environment of the deeply embedded protostars on scales from 400 to 8\,000~AU. The main conclusions of the paper are:

   \begin{enumerate}

      \item In the combined SMA/APEX data, we find two extended ridges of luminous H$_2$CO emission. They are not associated with the YSOs previously observed in the IRS7 region. The H$_2$CO emission extends across the whole field-of-view of the observations, spanning more than $10\,000$~AU.
      \item The physical temperatures in the two ridges around IRS7 were estimated with both LTE and non-LTE methods. They were in both cases found to be no lower than 30~K, more likely 50--60~K. These temperatures are too high to be attributed to internal heating from the low-mass YSOs in the region. The peculiar shape of the ridges, and their location with respect to the YSOs, also suggest that this region is strongly affected by external heating.
      \item The primary suspect for the external heating of the ridges is R~CrA, a $3~M_{\sun}$ Herbig Be star in the northwestern parts of the cloud. The morphology of the H$_2$CO-luminous ridges is relatively symmetric around the YSOs, which could indicate that the ridges are parts of a spherical shell around the low-mass YSOs.
      \item It was possible through radiative transfer modelling to explain the elevated temperatures by the UV radiative contribution from R~CrA. As discussed in Sect.~\ref{sec:intro}, earlier studies have shown that the evolution of a protostar can be affected by high-mass stars in their environments, but this study suggests that also intermediate-mass stars can have a noticeable effect on the physical properties in a whole protostellar envelope. The model also suggests an internal luminosity of $4~L_{\sun}$ and an envelope mass of $2.2~M_{\sun}$ for R~CrA IRS7B.
      \item The higher temperature of the southern ridge as compared to that of the northern ridge suggests that the southern ridge is actually closer to R~CrA than the northern ridge is, as a projection effect of the ridges being situated in different planes of sky.
      \item The strong UV flux from R~CrA obviously has an impact on the gas-phase chemistry on large scales in the region. Potentially it could also be critical for the ice chemistry, either inhibiting CO freeze-out or promoting formation of complex organics. The effect on the chemistry depends on how early in the evolution of the protostellar envelope around IRS7B the luminosity from R~CrA started to dominate the region.
      The elevated temperature in the cloud could also explain the unusual chemistry in the region, since much of the CO would evaporate from the dust grains at a very early stage, which would make the formation of complex organics on the dust grains inhibited.
   \end{enumerate}

This paper demonstrates the importance of understanding the nearby environment in the studies of protostars, and also the need for high-resolution observations. Unresolved single-dish observations of regions similar to IRS7 would not have shown the large-structure high-temperature gas found in this study, and without the knowledge of these structures, a high H$_2$CO temperature would likely have been attributed to the central source. Single-dish observations might still be useful in the comparison of the excitation of different molecules given the prior knowledge from this work. Thus, we suggest observations of more molecular lines to establish whether the high rotational temperature of H$_2$CO is reflected also in other species.

With the advent of ALMA, even more high-resolved studies of protostars will be made possible, making investigations where heating of the envelope from the central source and external stars are possible to distinguish. To study the more central and warmer parts of the protostellar envelopes, observations of higher frequency are needed. Such observations are currently being provided by \textit{Herschel Space Observatory}, and will be discussed in a future paper (Lindberg et~al., in prep.). These observations will make it possible to study whether the high level of excitation of the molecular gas is mainly caused by a PDR or by shocks.

\begin{acknowledgements}
We would like to thank the anonymous referee for comments and suggestions, which helped improving the manuscript. This research was supported by a grant from the Instrument Center for Danish Astrophysics (IDA) and a Lundbeck Foundation Group Leader Fellowship to JKJ. Research at Centre for Star and Planet Formation is funded by the Danish National Research Foundation and the University of Copenhagen's programme of excellence.

\end{acknowledgements}

\bibliographystyle{aa} 
\bibliography{h2co_rcra_final}

\begin{thebibliography}{59}
\expandafter\ifx\csname natexlab\endcsname\relax\def\natexlab#1{#1}\fi

\bibitem[{{Andr{\'e}} {et~al.}(1993){Andr{\'e}}, {Ward-Thompson}, \&
  {Barsony}}]{andre93}
{Andr{\'e}}, P., {Ward-Thompson}, D., \& {Barsony}, M. 1993, \apj, 406, 122

\bibitem[{{Andr{\'e}} {et~al.}(2000){Andr{\'e}}, {Ward-Thompson}, \&
  {Barsony}}]{andre00}
{Andr{\'e}}, P., {Ward-Thompson}, D., \& {Barsony}, M. 2000, Protostars and
  Planets IV, 59

\bibitem[{{Bergin} \& {Tafalla}(2007)}]{bergin07}
{Bergin}, E.~A. \& {Tafalla}, M. 2007, \araa, 45, 339

\bibitem[{{Bibo} {et~al.}(1992){Bibo}, {Th{\'e}}, \& {Dawanas}}]{bibo92}
{Bibo}, E.~A., {Th{\'e}}, P.~S., \& {Dawanas}, D.~N. 1992, \aap, 260, 293

\bibitem[{{Bottinelli} {et~al.}(2004{\natexlab{a}}){Bottinelli}, {Ceccarelli},
  {Lefloch}, {Williams}, {Castets}, {Caux}, {Cazaux}, {Maret}, {Parise}, \&
  {Tielens}}]{bottinelli04a}
{Bottinelli}, S., {Ceccarelli}, C., {Lefloch}, B., {et~al.} 2004{\natexlab{a}},
  \apj, 615, 354

\bibitem[{{Bottinelli} {et~al.}(2004{\natexlab{b}}){Bottinelli}, {Ceccarelli},
  {Neri}, {Williams}, {Caux}, {Cazaux}, {Lefloch}, {Maret}, \&
  {Tielens}}]{bottinelli04b}
{Bottinelli}, S., {Ceccarelli}, C., {Neri}, R., {et~al.} 2004{\natexlab{b}},
  \apjl, 617, L69

\bibitem[{{Brown}(1987)}]{brown87}
{Brown}, A. 1987, \apjl, 322, L31

\bibitem[{{Ceccarelli} {et~al.}(2007){Ceccarelli}, {Caselli}, {Herbst},
  {Tielens}, \& {Caux}}]{ceccarelli07}
{Ceccarelli}, C., {Caselli}, P., {Herbst}, E., {Tielens}, A.~G.~G.~M., \&
  {Caux}, E. 2007, Protostars and Planets V, 47

\bibitem[{{Chen} \& {Arce}(2010)}]{chen10}
{Chen}, X. \& {Arce}, H.~G. 2010, \apjl, 720, L169

\bibitem[{{Choi} {et~al.}(2008){Choi}, {Hamaguchi}, {Lee}, \&
  {Tatematsu}}]{choi08}
{Choi}, M., {Hamaguchi}, K., {Lee}, J., \& {Tatematsu}, K. 2008, \apj, 687, 406

\bibitem[{{Di Francesco} {et~al.}(2008){Di Francesco}, {Johnstone}, {Kirk},
  {MacKenzie}, \& {Ledwosinska}}]{difrancesco08}
{Di Francesco}, J., {Johnstone}, D., {Kirk}, H., {MacKenzie}, T., \&
  {Ledwosinska}, E. 2008, \apjs, 175, 277

\bibitem[{{Dickens} \& {Irvine}(1999)}]{dickens99}
{Dickens}, J.~E. \& {Irvine}, W.~M. 1999, \apj, 518, 733

\bibitem[{{Dullemond} {et~al.}(2002){Dullemond}, {van Zadelhoff}, \&
  {Natta}}]{transphere}
{Dullemond}, C.~P., {van Zadelhoff}, G.~J., \& {Natta}, A. 2002, \aap, 389, 464

\bibitem[{{Evans}(1999)}]{evans99}
{Evans}, II, N.~J. 1999, \araa, 37, 311

\bibitem[{{Forbrich} {et~al.}(2006){Forbrich}, {Preibisch}, \&
  {Menten}}]{forbrich06}
{Forbrich}, J., {Preibisch}, T., \& {Menten}, K.~M. 2006, \aap, 446, 155

\bibitem[{{Fuente} {et~al.}(1995){Fuente}, {Mart{\'i}n-Pintado}, \&
  {Gaume}}]{fuente95}
{Fuente}, A., {Mart{\'i}n-Pintado}, J., \& {Gaume}, R. 1995, \apjl, 442, L33

\bibitem[{{Goldsmith} \& {Langer}(1999)}]{goldsmith99}
{Goldsmith}, P.~F. \& {Langer}, W.~D. 1999, \apj, 517, 209

\bibitem[{{Gray} {et~al.}(2006){Gray}, {Corbally}, {Garrison}, {McFadden},
  {Bubar}, {McGahee}, {O'Donoghue}, \& {Knox}}]{gray06}
{Gray}, R.~O., {Corbally}, C.~J., {Garrison}, R.~F., {et~al.} 2006, \aj, 132,
  161

\bibitem[{{Green}(1991)}]{green91}
{Green}, S. 1991, \apjs, 76, 979

\bibitem[{{Groppi} {et~al.}(2007){Groppi}, {Hunter}, {Blundell}, \&
  {Sandell}}]{groppi07}
{Groppi}, C.~E., {Hunter}, T.~R., {Blundell}, R., \& {Sandell}, G. 2007, \apj,
  670, 489

\bibitem[{{Harju} {et~al.}(1993){Harju}, {Haikala}, {Mattila}, {Mauersberger},
  {Booth}, \& {Nordh}}]{harju93}
{Harju}, J., {Haikala}, L.~K., {Mattila}, K., {et~al.} 1993, \aap, 278, 569

\bibitem[{{Hartigan} \& {Graham}(1987)}]{hartigan87}
{Hartigan}, P. \& {Graham}, J.~A. 1987, \aj, 93, 913

\bibitem[{{Herbst} \& {van Dishoeck}(2009)}]{herbst09}
{Herbst}, E. \& {van Dishoeck}, E.~F. 2009, \araa, 47, 427

\bibitem[{{Ho} {et~al.}(2004){Ho}, {Moran}, \& {Lo}}]{ho04}
{Ho}, P.~T.~P., {Moran}, J.~M., \& {Lo}, K.~Y. 2004, \apjl, 616, L1

\bibitem[{{Hogerheijde} \& {van der Tak}(2000)}]{ratran}
{Hogerheijde}, M.~R. \& {van der Tak}, F.~F.~S. 2000, \aap, 362, 697

\bibitem[{{Jansen} {et~al.}(1995){Jansen}, {Spaans}, {Hogerheijde}, \& {van
  Dishoeck}}]{jansen95}
{Jansen}, D.~J., {Spaans}, M., {Hogerheijde}, M.~R., \& {van Dishoeck}, E.~F.
  1995, \aap, 303, 541

\bibitem[{{Johnstone} {et~al.}(1998){Johnstone}, {Hollenbach}, \&
  {Bally}}]{johnstone98}
{Johnstone}, D., {Hollenbach}, D., \& {Bally}, J. 1998, \apj, 499, 758

\bibitem[{{J{\o}rgensen}(2004)}]{jorgensen04_l483}
{J{\o}rgensen}, J.~K. 2004, \aap, 424, 589

\bibitem[{{J{\o}rgensen} {et~al.}(2007){J{\o}rgensen}, {Bourke}, {Myers}, {Di
  Francesco}, {van Dishoeck}, {Lee}, {Ohashi}, {Sch{\"o}ier}, {Takakuwa},
  {Wilner}, \& {Zhang}}]{jorgensen07}
{J{\o}rgensen}, J.~K., {Bourke}, T.~L., {Myers}, P.~C., {et~al.} 2007, \apj,
  659, 479

\bibitem[{{J{\o}rgensen} {et~al.}(2006){J{\o}rgensen}, {Johnstone}, {van
  Dishoeck}, \& {Doty}}]{jorgensen06}
{J{\o}rgensen}, J.~K., {Johnstone}, D., {van Dishoeck}, E.~F., \& {Doty}, S.~D.
  2006, \aap, 449, 609

\bibitem[{{J{\o}rgensen} {et~al.}(2005){J{\o}rgensen}, {Sch{\"o}ier}, \& {van
  Dishoeck}}]{jorgensen05}
{J{\o}rgensen}, J.~K., {Sch{\"o}ier}, F.~L., \& {van Dishoeck}, E.~F. 2005,
  \aap, 437, 501

\bibitem[{{Mangum} \& {Wootten}(1993)}]{mangum93}
{Mangum}, J.~G. \& {Wootten}, A. 1993, \apjs, 89, 123

\bibitem[{{Maret} {et~al.}(2004){Maret}, {Ceccarelli}, {Caux}, {Tielens},
  {J{\o}rgensen}, {van Dishoeck}, {Bacmann}, {Castets}, {Lefloch}, {Loinard},
  {Parise}, \& {Sch{\"o}ier}}]{maret04}
{Maret}, S., {Ceccarelli}, C., {Caux}, E., {et~al.} 2004, \aap, 416, 577

\bibitem[{{Miettinen} {et~al.}(2008){Miettinen}, {Kontinen}, {Harju}, \&
  {Higdon}}]{miettinen08}
{Miettinen}, O., {Kontinen}, S., {Harju}, J., \& {Higdon}, J.~L. 2008, \aap,
  486, 799

\bibitem[{{M{\"u}ller} {et~al.}(2001){M{\"u}ller}, {Thorwirth}, {Roth}, \&
  {Winnewisser}}]{cdms}
{M{\"u}ller}, H.~S.~P., {Thorwirth}, S., {Roth}, D.~A., \& {Winnewisser}, G.
  2001, \aap, 370, L49

\bibitem[{{Neuh{\"a}user} \& {Forbrich}(2008)}]{neuhauser08}
{Neuh{\"a}user}, R. \& {Forbrich}, J. 2008, in Handbook of Star Forming
  Regions, Volume II: The Southern Sky, ed. {Reipurth, B.} (ASP Monographs; San
  Francisco, CA; ASP), 735

\bibitem[{{Nutter} {et~al.}(2005){Nutter}, {Ward-Thompson}, \&
  {Andr{\'e}}}]{nutter05}
{Nutter}, D.~J., {Ward-Thompson}, D., \& {Andr{\'e}}, P. 2005, \mnras, 357, 975

\bibitem[{{{\"O}berg} {et~al.}(2009){{\"O}berg}, {Garrod}, {van Dishoeck}, \&
  {Linnartz}}]{oberg09}
{{\"O}berg}, K.~I., {Garrod}, R.~T., {van Dishoeck}, E.~F., \& {Linnartz}, H.
  2009, \aap, 504, 891

\bibitem[{{O'Dell} {et~al.}(1993){O'Dell}, {Wen}, \& {Hu}}]{odell93}
{O'Dell}, C.~R., {Wen}, Z., \& {Hu}, X. 1993, \apj, 410, 696

\bibitem[{{Ossenkopf} \& {Henning}(1994)}]{ossenkopf94}
{Ossenkopf}, V. \& {Henning}, T. 1994, \aap, 291, 943

\bibitem[{{Peterson} {et~al.}(2011){Peterson}, {Caratti o Garatti}, {Bourke},
  {Forbrich}, {Gutermuth}, {J{\o}rgensen}, {Allen}, {Patten}, {Dunham},
  {Harvey}, {Mer{\'{\i}}n}, {Chapman}, {Cieza}, {Huard}, {Knez}, {Prager}, \&
  {Evans}}]{peterson11}
{Peterson}, D.~E., {Caratti o Garatti}, A., {Bourke}, T.~L., {et~al.} 2011,
  \apjs, 194, 43

\bibitem[{{Pickett} {et~al.}(1998){Pickett}, {Poynter}, {Cohen}, {Delitsky},
  {Pearson}, \& {M{\"u}ller}}]{jpl}
{Pickett}, H.~M., {Poynter}, R.~L., {Cohen}, E.~A., {et~al.} 1998, \jqsrt, 60,
  883

\bibitem[{{Qi}(2012)}]{qimir}
{Qi}, C. 2012, The MIR Cookbook, The Submillimeter Array / Harvard-Smithsonian
  Center for Astrophysics (http://cfa-www.harvard.edu/$\sim$cqi/mircook.html)

\bibitem[{{Sakai} {et~al.}(2009{\natexlab{a}}){Sakai}, {Sakai}, {Hirota},
  {Burton}, \& {Yamamoto}}]{sakai09a}
{Sakai}, N., {Sakai}, T., {Hirota}, T., {Burton}, M., \& {Yamamoto}, S.
  2009{\natexlab{a}}, \apj, 697, 769

\bibitem[{{Sakai} {et~al.}(2009{\natexlab{b}}){Sakai}, {Sakai}, {Hirota}, \&
  {Yamamoto}}]{sakai09b}
{Sakai}, N., {Sakai}, T., {Hirota}, T., \& {Yamamoto}, S. 2009{\natexlab{b}},
  \apj, 702, 1025

\bibitem[{{Sault} {et~al.}(1995){Sault}, {Teuben}, \& {Wright}}]{sault95}
{Sault}, R.~J., {Teuben}, P.~J., \& {Wright}, M.~C.~H. 1995, in Astronomical
  Society of the Pacific Conference Series, Vol.~77, Astronomical Data Analysis
  Software and Systems IV, ed. {R.~A.~Shaw, H.~E.~Payne, \& J.~J.~E.~Hayes},
  433

\bibitem[{{Sch{\"o}ier} {et~al.}(2006){Sch{\"o}ier}, {J{\o}rgensen},
  {Pontoppidan}, \& {Lundgren}}]{schoier06}
{Sch{\"o}ier}, F.~L., {J{\o}rgensen}, J.~K., {Pontoppidan}, K.~M., \&
  {Lundgren}, A.~A. 2006, \aap, 454, L67

\bibitem[{{Sch{\"o}ier} {et~al.}(2005){Sch{\"o}ier}, {van der Tak}, {van
  Dishoeck}, \& {Black}}]{lamda}
{Sch{\"o}ier}, F.~L., {van der Tak}, F.~F.~S., {van Dishoeck}, E.~F., \&
  {Black}, J.~H. 2005, \aap, 432, 369

\bibitem[{{Sternberg} \& {Dalgarno}(1995)}]{sternberg95}
{Sternberg}, A. \& {Dalgarno}, A. 1995, \apjs, 99, 565

\bibitem[{{Takakuwa} {et~al.}(2003){Takakuwa}, {Kamazaki}, {Saito}, \&
  {Hirano}}]{takakuwa03}
{Takakuwa}, S., {Kamazaki}, T., {Saito}, M., \& {Hirano}, N. 2003, \apj, 584,
  818

\bibitem[{{Taylor} \& {Storey}(1984)}]{taylor84}
{Taylor}, K.~N.~R. \& {Storey}, J.~W.~V. 1984, \mnras, 209, 5P

\bibitem[{{Tielens} \& {Hollenbach}(1985{\natexlab{a}})}]{tielens85a}
{Tielens}, A.~G.~G.~M. \& {Hollenbach}, D. 1985{\natexlab{a}}, \apj, 291, 747

\bibitem[{{Tielens} \& {Hollenbach}(1985{\natexlab{b}})}]{tielens85b}
{Tielens}, A.~G.~G.~M. \& {Hollenbach}, D. 1985{\natexlab{b}}, \apj, 291, 722

\bibitem[{{van der Tak} {et~al.}(2007){van der Tak}, {Black}, {Sch{\"o}ier},
  {Jansen}, \& {van Dishoeck}}]{vandertak07}
{van der Tak}, F.~F.~S., {Black}, J.~H., {Sch{\"o}ier}, F.~L., {Jansen}, D.~J.,
  \& {van Dishoeck}, E.~F. 2007, \aap, 468, 627

\bibitem[{{Wang} {et~al.}(2004){Wang}, {Mundt}, {Henning}, \& {Apai}}]{wang04}
{Wang}, H., {Mundt}, R., {Henning}, T., \& {Apai}, D. 2004, \apj, 617, 1191

\bibitem[{{Ward-Thompson} {et~al.}(2007){Ward-Thompson}, {Andr{\'e}},
  {Crutcher}, {Johnstone}, {Onishi}, \& {Wilson}}]{wardthompson07}
{Ward-Thompson}, D., {Andr{\'e}}, P., {Crutcher}, R., {et~al.} 2007, Protostars
  and Planets V, 33

\bibitem[{{Watanabe} {et~al.}(2012){Watanabe}, {Sakai}, {Lindberg},
  {J{\o}rgensen}, {Bisschop}, \& {Yamamoto}}]{watanabe12}
{Watanabe}, Y., {Sakai}, N., {Lindberg}, J.~E., {et~al.} 2012, \apj, 745, 126

\bibitem[{{Wei{\ss}} {et~al.}(2001){Wei{\ss}}, {Neininger}, {H{\"u}ttemeister},
  \& {Klein}}]{weiss01}
{Wei{\ss}}, A., {Neininger}, N., {H{\"u}ttemeister}, S., \& {Klein}, U. 2001,
  \aap, 365, 571

\bibitem[{{Wilson} \& {Rood}(1994)}]{wilson94}
{Wilson}, T.~L. \& {Rood}, R. 1994, \araa, 32, 191

\end{thebibliography}

\begin{appendix}

\section{Spectral line channel maps}

\begin{figure*}[!htb]
    \centering
    \includegraphics{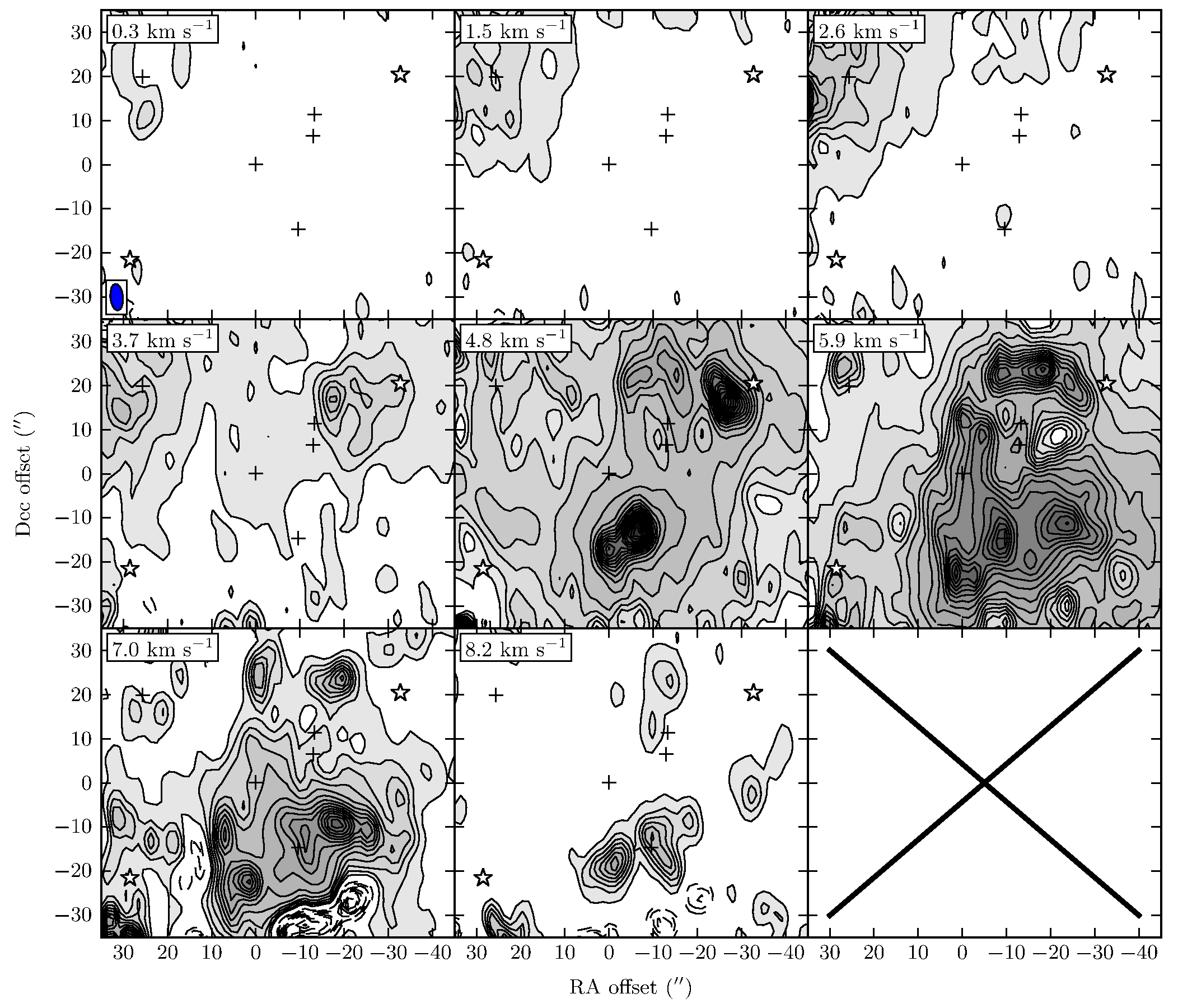}
    \caption{SMA/APEX channel maps of the H$_2$CO 3$_{03}$\nobrl 2$_{02}$ line. Contours every 0.5~\jybeam~\kms\ ($\sim3\sigma$ at edge of primary beam). Refer to Fig.\ref{fig:herschelsmaspitzer} for a guide to the symbols used for the compact objects.}\label{fig:h2co1}
\end{figure*}

\newpage

\begin{figure*}[!htb]
    \centering
    \includegraphics{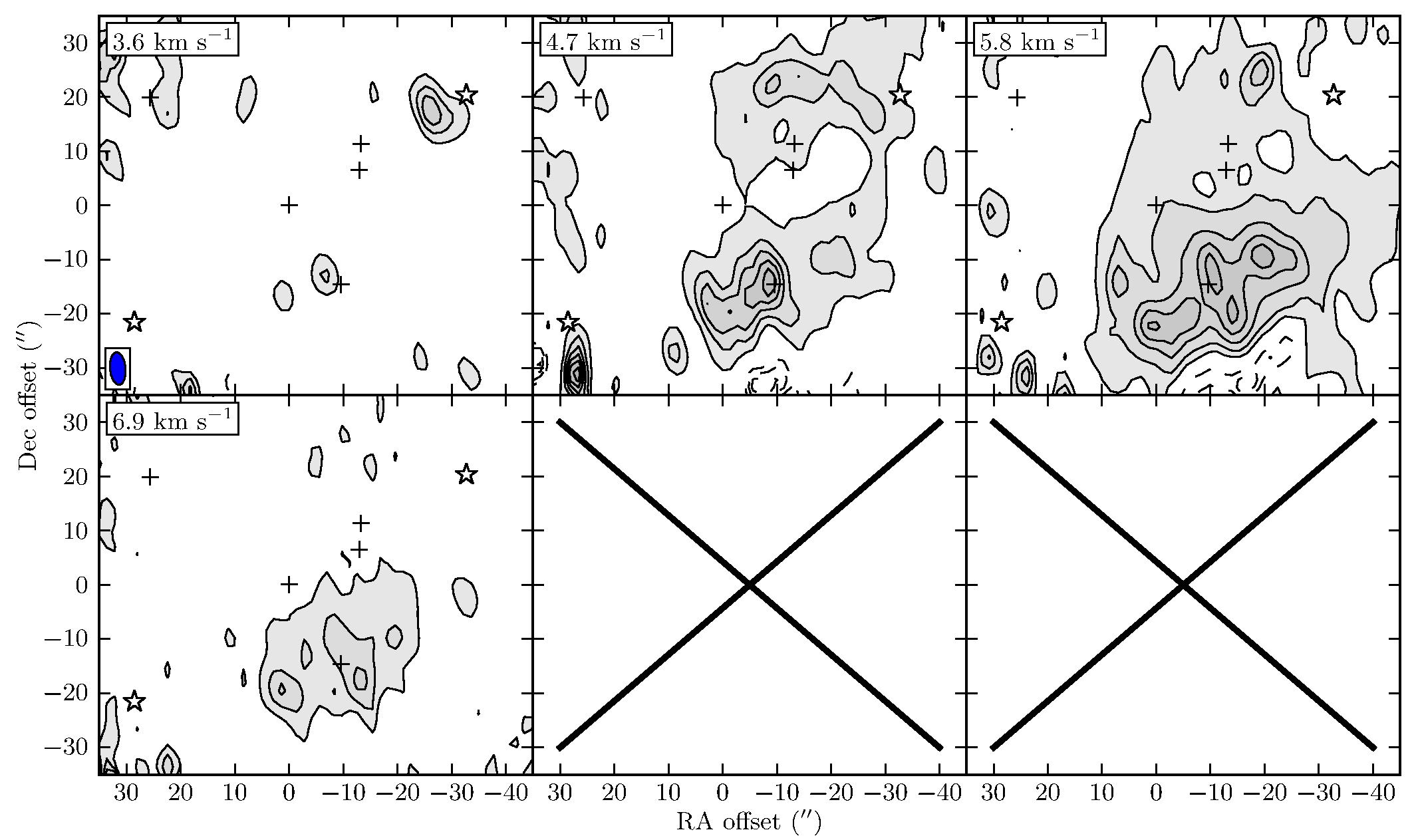}
    \caption{SMA/APEX channel maps of the H$_2$CO 3$_{21}$\nobrl 2$_{20}$ line. Contours every 0.5~\jybeam~\kms\ ($\sim3\sigma$ at edge of primary beam). Refer to Fig.\ref{fig:herschelsmaspitzer} for a guide to the symbols used for the compact objects.}\label{fig:h2co3}
\end{figure*}

\newpage	

\begin{figure*}[!htb]
    \centering
    \includegraphics{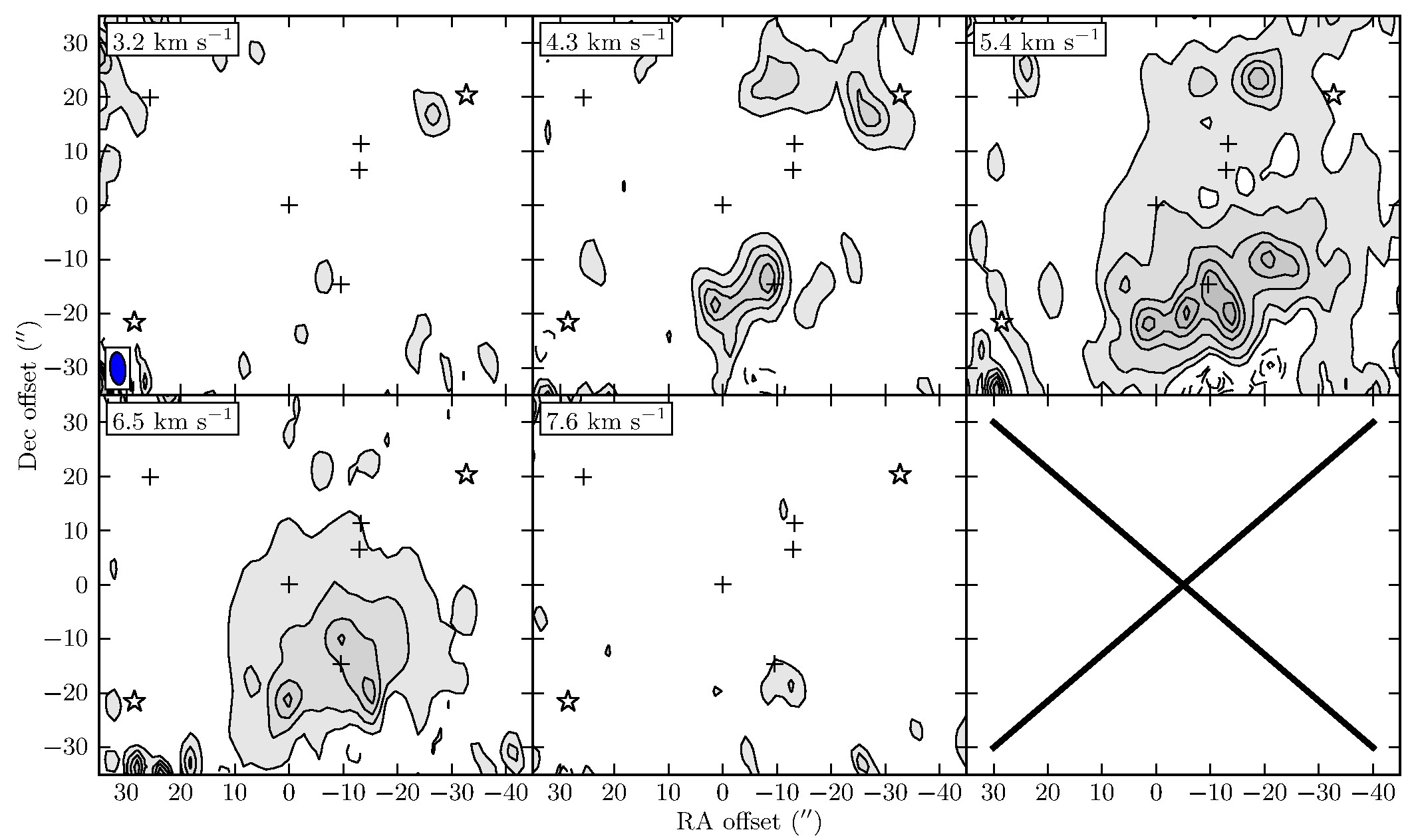}
    \caption{SMA/APEX channel maps of the H$_2$CO 3$_{22}$\nobrl 2$_{21}$ line. Contours every 0.5~\jybeam~\kms\ ($\sim3\sigma$ at edge of primary beam). Refer to Fig.\ref{fig:herschelsmaspitzer} for a guide to the symbols used for the compact objects.}\label{fig:h2co4}
\end{figure*}

\begin{figure*}[!htb]
    \centering
    \includegraphics{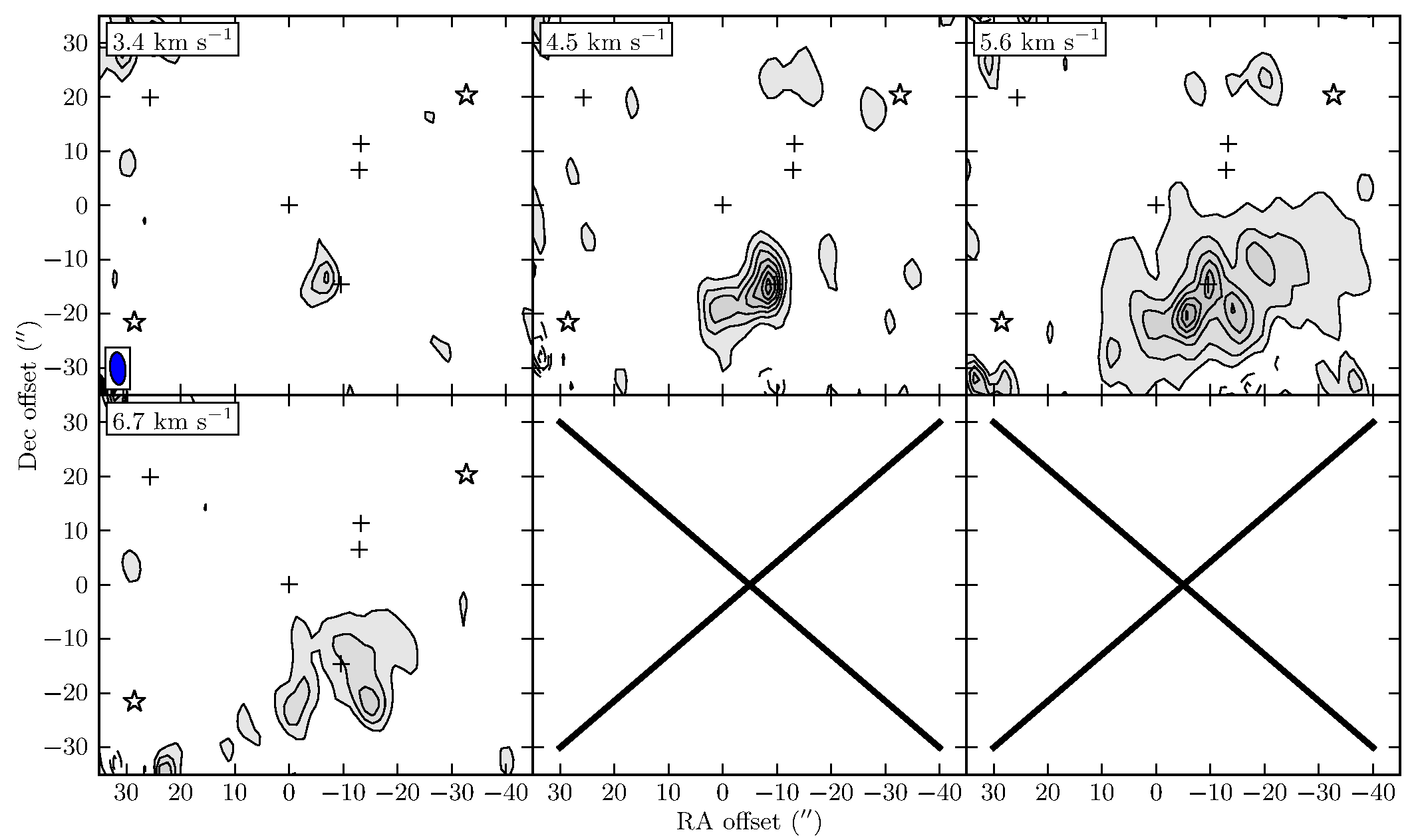}
    \caption{SMA/APEX channel maps of the CH$_3$OH 4$_{22}$\nobrl 3$_{12}$ line. Contours every 0.5~\jybeam~\kms\ ($\sim3\sigma$ at edge of primary beam). Refer to Fig.\ref{fig:herschelsmaspitzer} for a guide to the symbols used for the compact objects.}\label{fig:ch3oh}
\end{figure*}

\newpage

\begin{figure*}[!htb]
    \centering
    \includegraphics{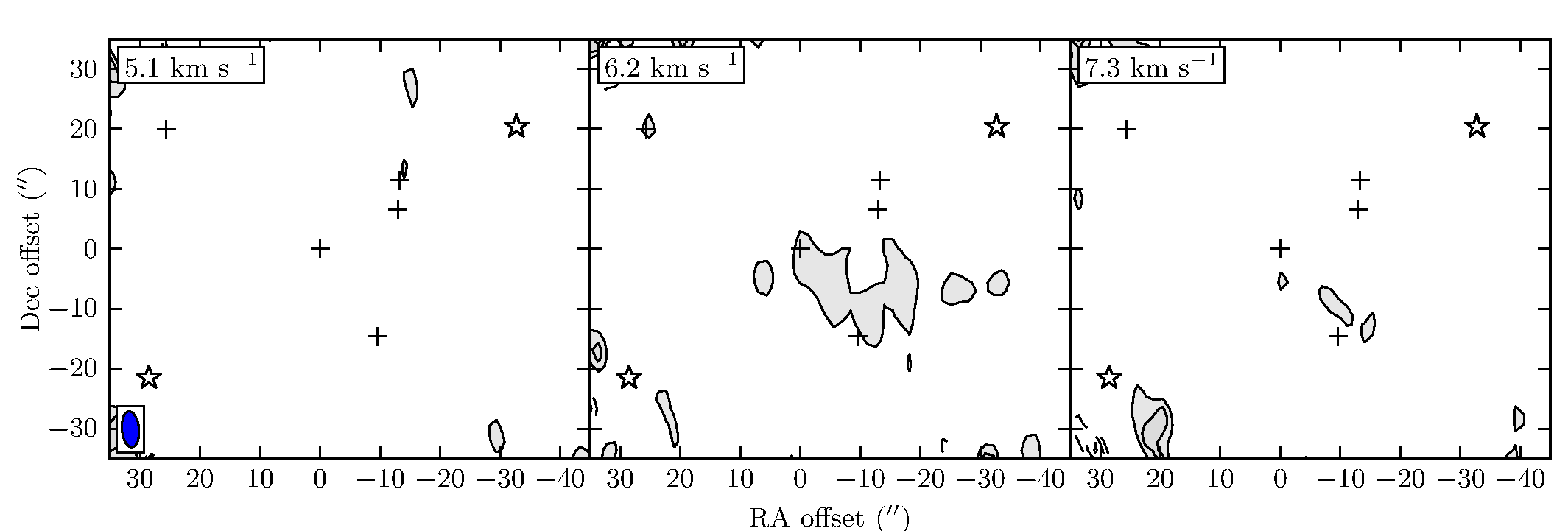}
    \caption{SMA/APEX channel maps of the HC$_3$N 24\nobrl 23 line. Contours every 0.5~\jybeam~\kms\ ($\sim3\sigma$ at edge of primary beam). Refer to Fig.\ref{fig:herschelsmaspitzer} for a guide to the symbols used for the compact objects.}\label{fig:hc3n}
\end{figure*}

\begin{figure*}[!htb]
    \centering
    \includegraphics{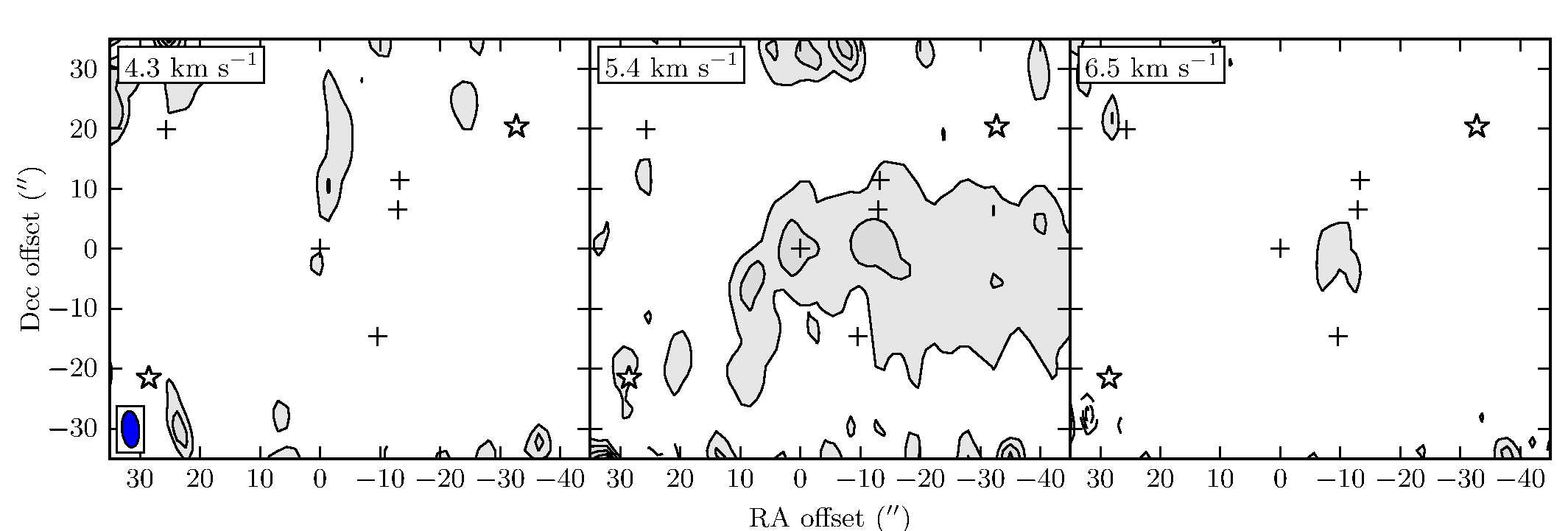}
    \caption{SMA/APEX channel maps of the \textit{c}\nobr C$_3$H$_2$ 6$_{16}$\nobrl 5$_{05}$ and 6$_{06}$\nobrl 5$_{15}$ lines (blended). Contours every 0.5~\jybeam~\kms\ ($\sim3\sigma$ at edge of primary beam). Refer to Fig.\ref{fig:herschelsmaspitzer} for a guide to the symbols used for the compact objects.}\label{fig:c3h21}
\end{figure*}

\begin{figure*}[!htb]
    \centering
    \includegraphics{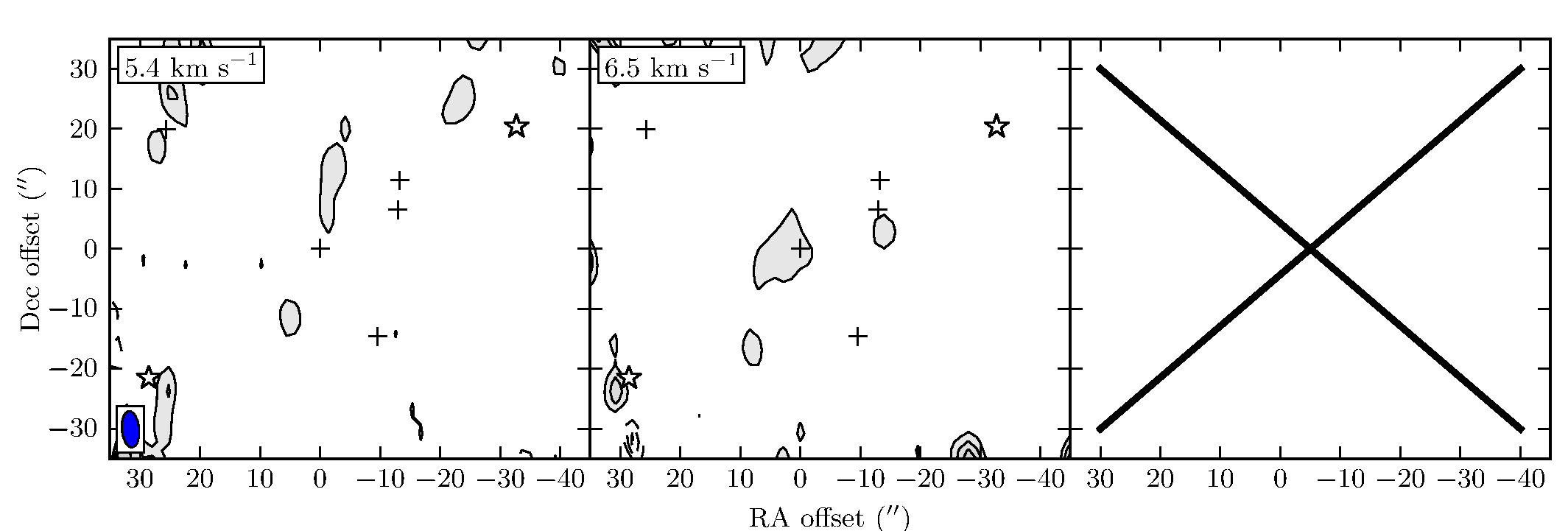}
    \caption{SMA/APEX channel maps of the \textit{c}\nobr C$_3$H$_2$ 5$_{14}$\nobrl 4$_{23}$ line. Contours every 0.5~\jybeam~\kms\ ($\sim3\sigma$ at edge of primary beam). Refer to Fig.\ref{fig:herschelsmaspitzer} for a guide to the symbols used for the compact objects.}\label{fig:c3h22}
\end{figure*}

\newpage

\begin{figure*}[!htb]
    \centering
    \includegraphics{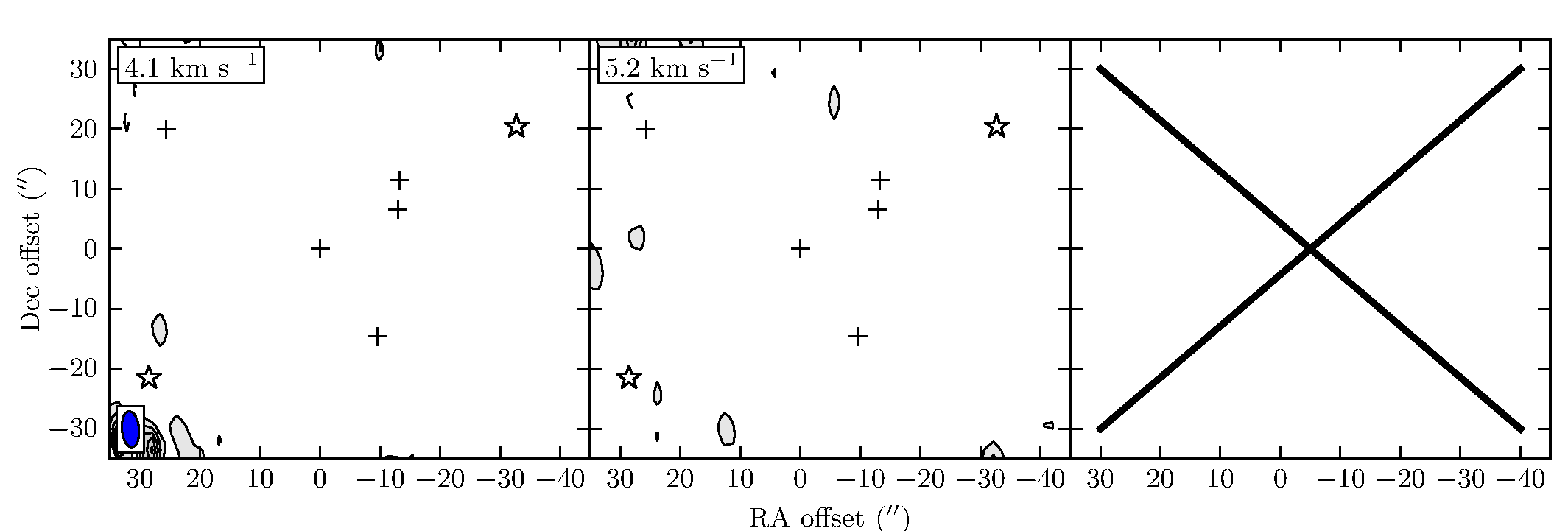}
    \caption{SMA/APEX channel maps of the \textit{c}\nobr C$_3$H$_2$ 5$_{24}$\nobrl 4$_{13}$ line. Contours every 0.5~\jybeam~\kms\ ($\sim3\sigma$ at edge of primary beam). Refer to Fig.\ref{fig:herschelsmaspitzer} for a guide to the symbols used for the compact objects.}\label{fig:c3h23}
\end{figure*}

\begin{figure*}[!htb]
    \centering
    \includegraphics{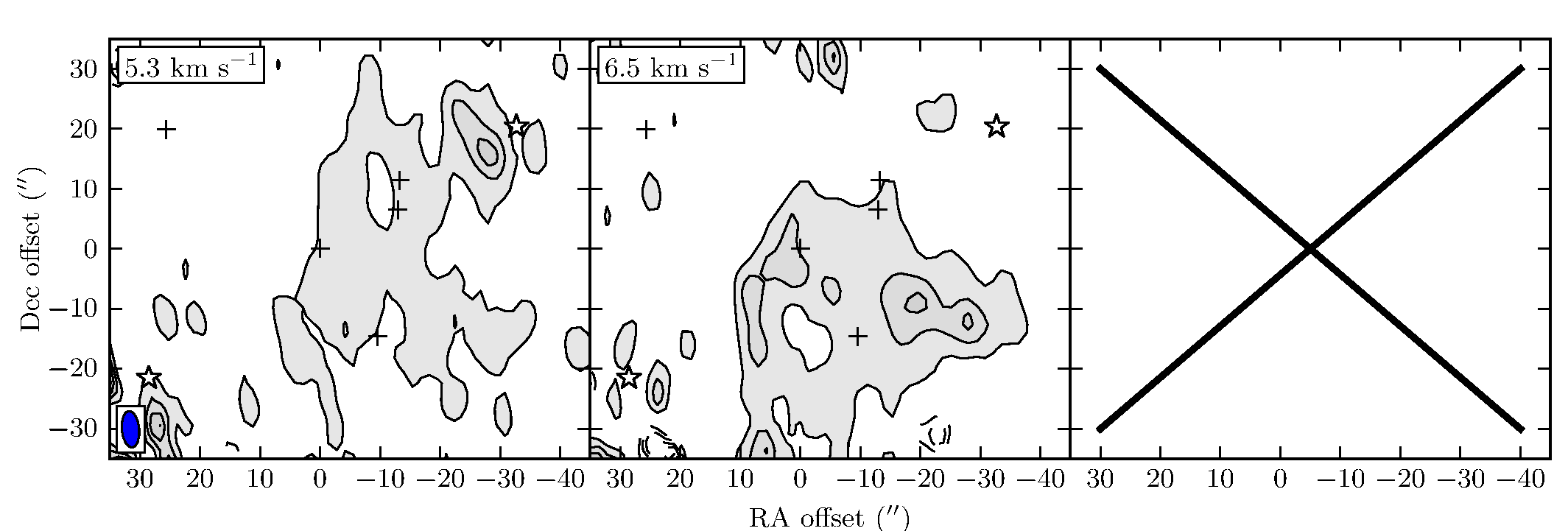}
    \caption{SMA/APEX channel maps of the DCN 3\nobrl 2 line. Contours every 0.5~\jybeam~\kms\ ($\sim3\sigma$ at edge of primary beam). Refer to Fig.\ref{fig:herschelsmaspitzer} for a guide to the symbols used for the compact objects.}\label{fig:dcn}
\end{figure*}

\begin{figure*}[!htb]
    \centering
    \includegraphics{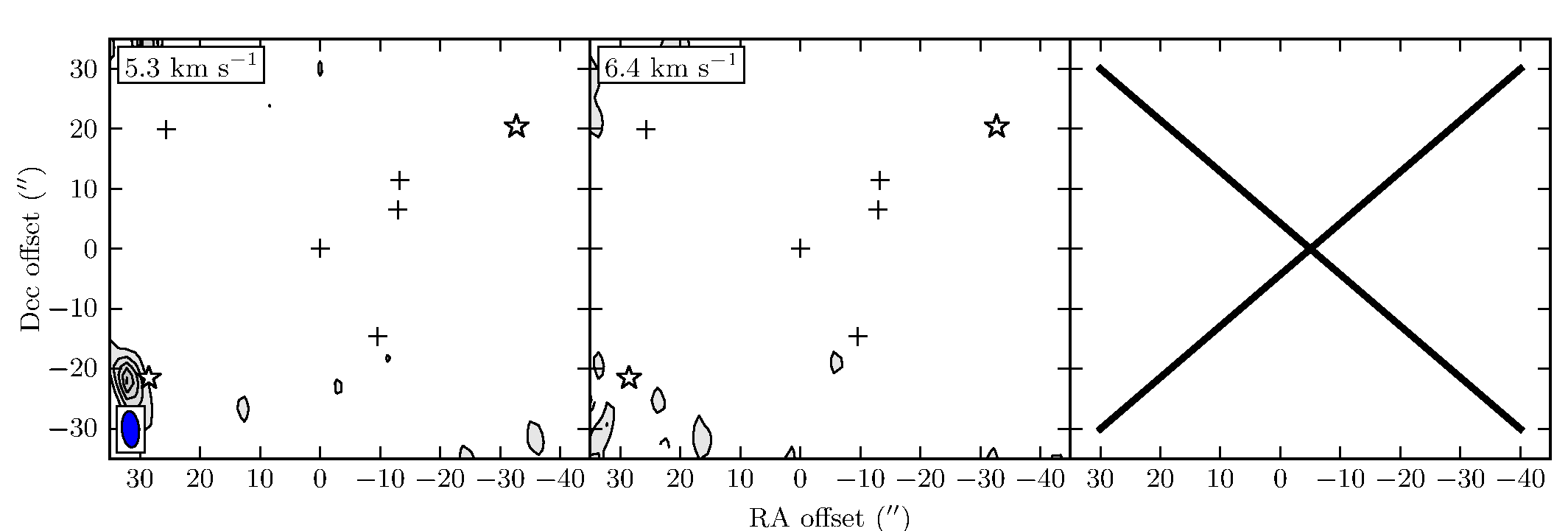}
    \caption{SMA/APEX channel maps of the $^{13}$CN $N$ = 2\nobrl 1, $J$ = 5/2\nobrl 3/2 $F_1$ = 3\nobrl 2, $F$ = 4\nobrl 3; $F_1$ = 3\nobrl 2, $F$ = 3\nobrl 2; and  $F_1$ = 3\nobrl 2, $F$ = 2\nobrl 1 lines. Contours every 0.5~\jybeam~\kms\ ($\sim3\sigma$ at edge of primary beam). Refer to Fig.\ref{fig:herschelsmaspitzer} for a guide to the symbols used for the compact objects.}\label{fig:13cna}
\end{figure*}

\newpage

\begin{figure*}[!htb]
    \centering
    \includegraphics{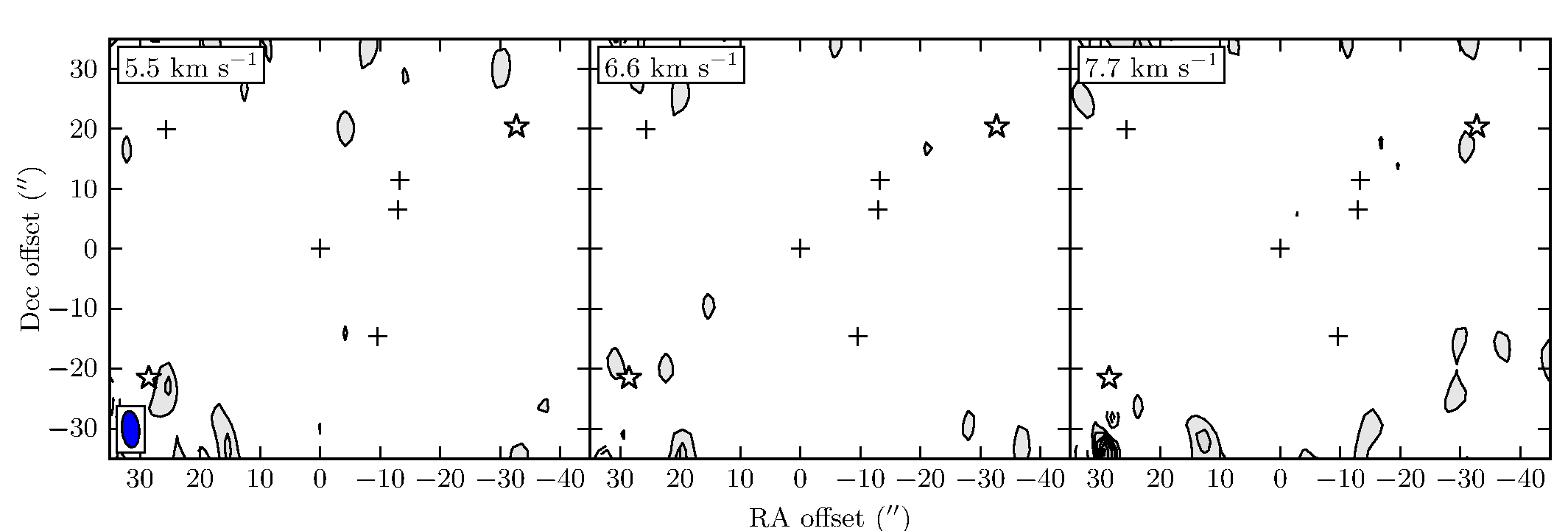}
    \caption{SMA/APEX channel maps of the $^{13}$CN $N$ = 2\nobrl 1, $J$ = 3/2\nobrl 1/2 $F_1$ = 1\nobrl 0, $F$ = 0\nobrl 1 line. Contours every 0.5~\jybeam~\kms\ ($\sim3\sigma$ at edge of primary beam). Refer to Fig.\ref{fig:herschelsmaspitzer} for a guide to the symbols used for the compact objects.}\label{fig:13cnb}
\end{figure*}

\begin{figure*}[!htb]
    \centering
    \includegraphics{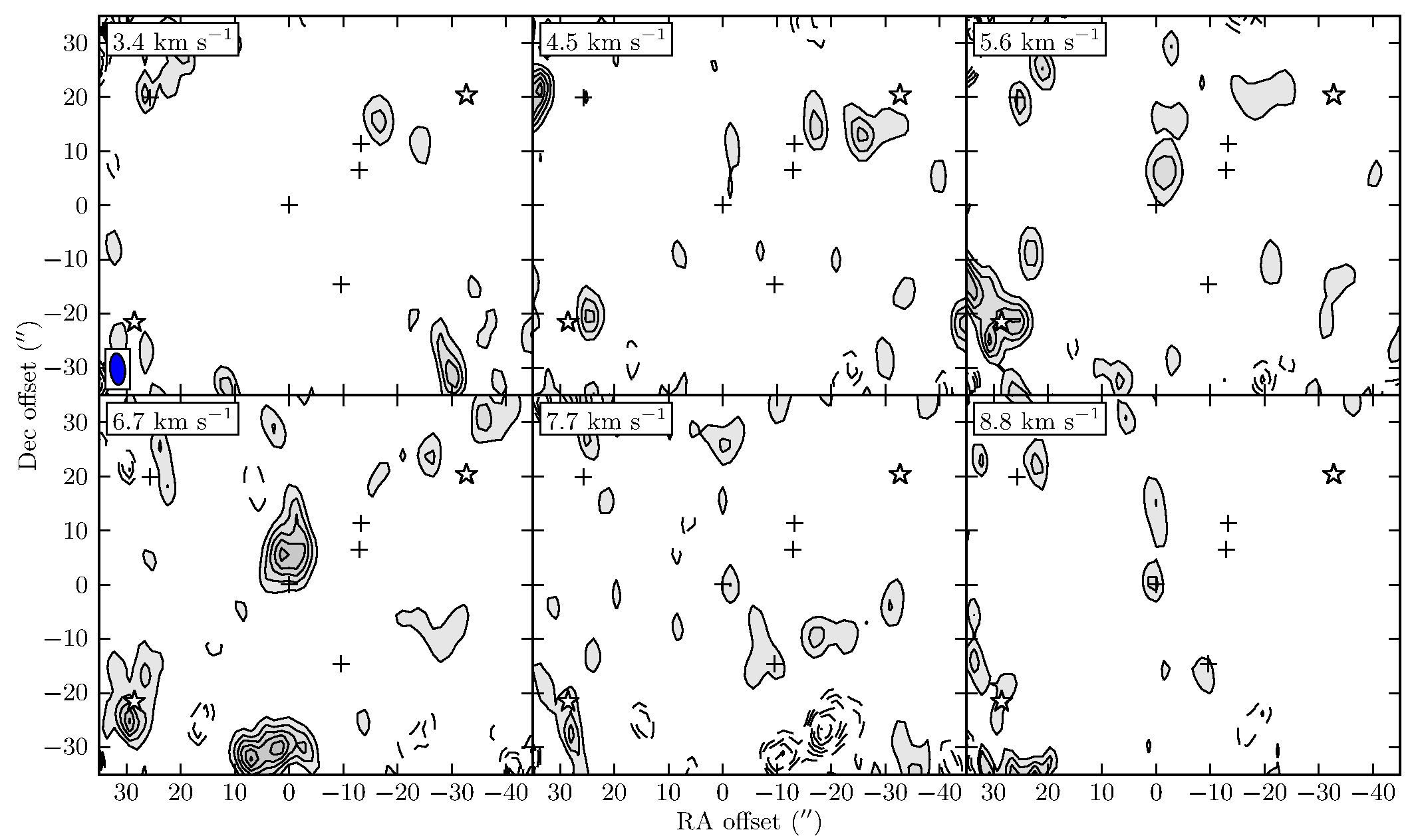}
    \caption{SMA channel maps of the $^{12}$CN 	N= 2\nobrl 1, J=5/2\nobrl 3/2 $F$ = 5/2\nobrl 3/2,  $F$ = 7/2\nobrl 5/2, and $F$ = 3/2\nobrl 1/2 lines. Contours every 0.5~\jybeam~\kms\ ($\sim3\sigma$ at edge of primary beam). No APEX short-spacing data were used for this line. Refer to Fig.\ref{fig:herschelsmaspitzer} for a guide to the symbols used for the compact objects.}\label{fig:12cn}
\end{figure*}

\newpage

\begin{figure*}[!htb]
    \centering
    \includegraphics{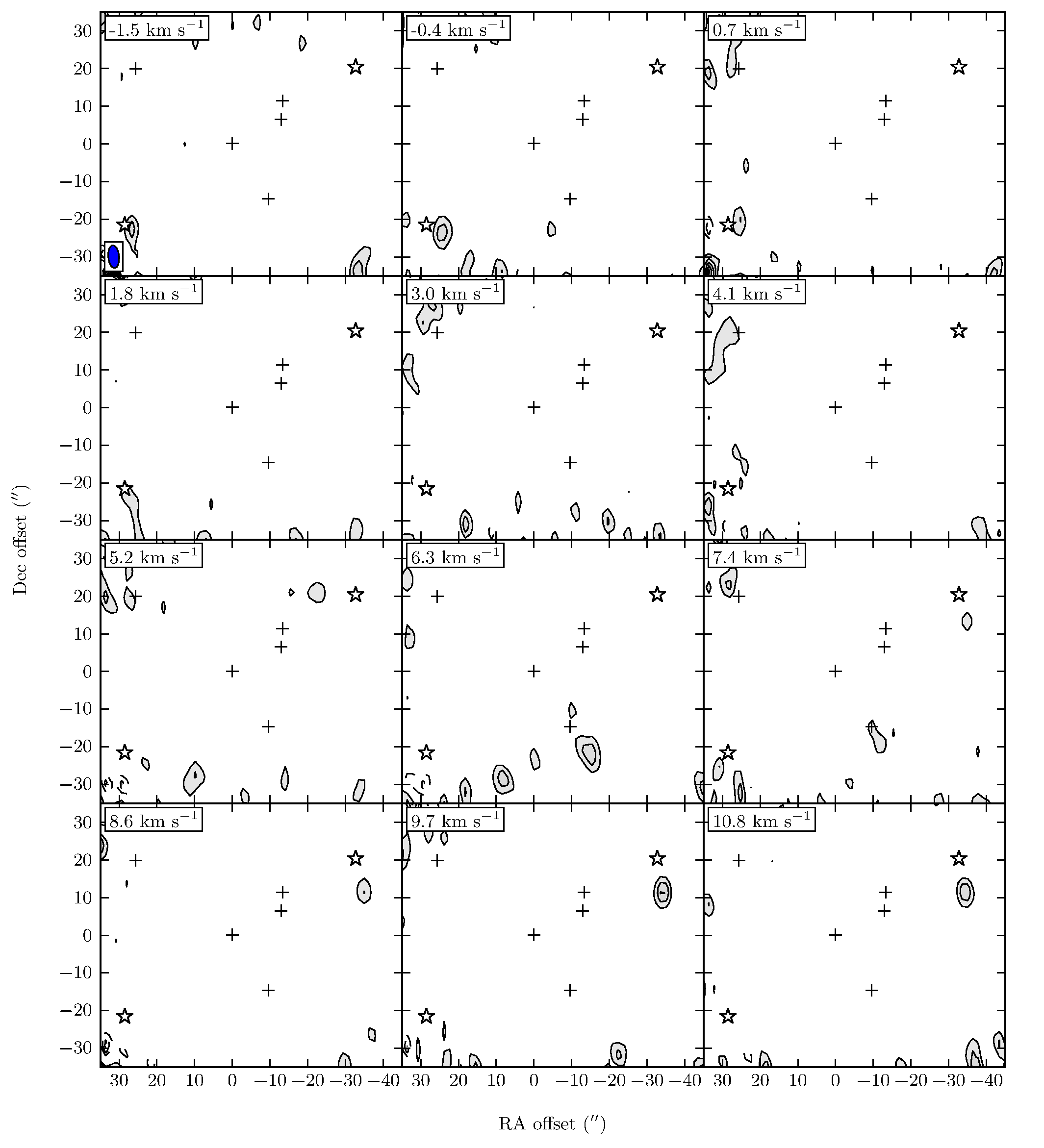}
    \caption{SMA/APEX channel maps of the SiO 5\nobrl 4 line. Contours every 0.5~\jybeam~\kms\ ($\sim3\sigma$ at edge of primary beam). Refer to Fig.\ref{fig:herschelsmaspitzer} for a guide to the symbols used for the compact objects.}\label{fig:sio}
\end{figure*}

\end{appendix}
\end{document}